\documentclass[prd,
,superscriptaddress,
nofootinbib,%
tightenlines ]{revtex4}
\usepackage{epsfig}
\usepackage[colorlinks=true,linkcolor=blue,urlcolor=blue,citecolor=blue]{hyperref}

\usepackage{amsfonts}
\usepackage{amssymb}
\usepackage{amsmath}
\usepackage{slashed}

\newcommand{\ben}{\begin{displaymath}}
\newcommand{\een}{\end{displaymath}}
\newcommand{\be}{\begin{equation}}
\newcommand{\ee}{\end{equation}}
\newcommand{\bea}{\begin{eqnarray}}
\newcommand{\eea}{\end{eqnarray}}

\usepackage[toc,page]{appendix}
\usepackage[ngerman, english]{babel}
\usepackage{pstricks}
\usepackage{color}
\usepackage{caption}
\usepackage{subcaption} 
\usepackage{float}
\usepackage{mwe}
\usepackage{enumitem}
\usepackage{cancel}

\usepackage{pdfpages}
\usepackage{hyperref}
\graphicspath{ {./images/} }

\begin{document}

\title{Reconstruction of gravitational form factors using generative machine learning}

\author{Herzallah~Alharazin}
\affiliation{Institut f\"ur Theoretische Physik II, Ruhr-Universit\"at Bochum, D-44780 Bochum, Germany}

\author{Julia~Yu.~Panteleeva}
\affiliation{Institut f\"ur Theoretische Physik II, Ruhr-Universit\"at Bochum, D-44780 Bochum, Germany}

\date{\today}

\begin{abstract}
We develop a generative framework based on denoising diffusion for the model-independent reconstruction of hadronic form factors from sparse and noisy data.~The generative prior is built from a large ensemble of synthetic curves drawn from ten distinct functional classes rooted in different theoretical approaches to hadron structure.~Applied to the proton gravitational form factors $A(t)$, $J(t)$, and $D(t)$, the framework yields non-parametric reconstructions consistent with lattice QCD across the full kinematic range $0\le -t\le 2~\mathrm{GeV}^{2}$, remaining robust even when only one or two conditioning points are retained.~The densely sampled output enables a direct extraction of the chiral low-energy constants $c_8=-4.6\pm 0.8\ \rm GeV^{-1}$ and $c_9=-0.61\pm 0.19\ \rm GeV^{-1}$.~Using these values at the physical pion mass, we obtain $D(0)=-4.3\pm 0.8$ for the nucleon $D$-term.
\end{abstract}

\maketitle

\section{Introduction}
\label{sec:introduction}

The hadronic matrix elements of the energy-momentum tensor (EMT) have been studied intensively over the past decades, both theoretically and experimentally.~While much attention has been devoted to the physical interpretation of the Fourier transforms of the gravitational form factors (GFFs) in terms of spatial density distributions of energy (mass), spin, pressure, and shear forces inside hadrons, see, e.g., \cite{Jaffe:2020ebz,Panteleeva:2022uii,Freese:2021mzg,Ji:2025qax}, the GFFs themselves, as functions of momentum transfer, are interesting in their own right.~In particular, they can be extracted from generalized parton distributions (GPDs)~\cite{Muller:1994ses,Ji:1996ek,Radyushkin:1996nd,Goeke:2001tz}, which are experimentally accessible through exclusive processes such as deeply virtual Compton scattering (DVCS)~\cite{Ji:1996ek,Radyushkin:1997ki, Burkert:2018bqq, Burkert:2021ith} and hard exclusive meson production~\cite{Collins:1996fb}.~Knowing the GFFs of a system that can be probed experimentally, such as the nucleon, establishes a direct connection between theoretical predictions and experiment, thereby testing the validity and consistency of the underlying theory.

On the theory side, there exist two main model-independent approaches to obtain the nucleon GFFs, each with its own limitations.~The first is chiral perturbation theory (ChPT), which is valid at low energies but suffers from the presence of unknown low-energy constants (LECs).~While it can predict the leading non-analytic behaviour of the GFFs at low energies, and hence the long-range behavior of their Fourier transforms in the chiral limit, it cannot provide definitive statements about the values of the GFFs near $t \approx 0$ \cite{Alharazin:2020yjv,Alharazin:2022wjj,Alharazin:2023uhr}.

The second approach is lattice QCD.~It performs well for the form factors $A(t)$ and $J(t)$, whose values at $t = 0$ are  constrained to $1$ and $1/2$, respectively, by the operator structure, and it covers a wide range of momentum transfer $t$.~However, current lattice QCD calculations of the GFFs  suffer from several systematic limitations:~most rely on a single gauge ensemble at a single lattice spacing and volume, preventing a controlled continuum and infinite-volume extrapolation.~Moreover, most calculations employ unphysical pion masses.~These limitations, combined with the intrinsically noisy signal, yield large uncertainties near $t = 0$  for $D(t)$, whose extrapolation from the smallest accessible  $|t|$ to $t=0$ introduces a model dependence~\cite{Shanahan:2018pib,Hackett:2023nkr}.

Unlike $A(t)$ and  $J(t)$, the value of $D(t)$ at $t = 0$ is not constrained by  the operator structure.~This value, the so-called $D$-term, represents a fundamental global property of the nucleon, considered as important as its mass and charge~\cite{Kobzarev:1962wt,Pagels:1966zza,Polyakov:2002yz,Polyakov:2018zvc}.~These limitations motivate the development of complementary frameworks that can provide reliable, parametrization-independent predictions for the full set of nucleon GFFs, in particular for the $D$-term.~Ideally, such a framework should possess two key properties.~First, it should be capable of producing robust reconstructions from a small number of precisely determined data points, making optimal use of their accuracy rather than relying on dense kinematic coverage.~As will be demonstrated in this work, the constraining power of a few well-placed, high-precision measurements can rival or exceed that of a larger but noisier dataset.~Second, it should be able to combine the sparse first-principles data from lattice QCD with the analytic constraints from ChPT, without committing to a specific functional form for the $t$-dependence.

Denoising diffusion probabilistic models (DDPMs)~\cite{Sohl-Dickstein:2015deep,Ho:2020ddpm,Song:2020score} have emerged as the leading class of deep generative models for high-resolution image synthesis~\cite{Rombach:2021ldi,Nichol:2021improved}, owing to their ability to learn complex, high-dimensional probability distributions and to generate diverse samples from conditional posteriors.~These same properties make them natural candidates for inverse problems in the natural sciences, where a small number of noisy observations must be completed into a full function or field configuration.~In fundamental physics, diffusion models have recently entered the domain of lattice field theory, where the formal connection between the denoising process and stochastic quantization has been established for scalar and Abelian gauge fields and subsequently exploited for the generation of non-Abelian lattice gauge configurations \cite{Wang:2023sqm,Zhu:2025pcdm,Aarts:2026zzr,Alharazin:2026su2}.

In this work, we develop a physics-guided DDPM framework for the model-independent reconstruction of hadronic form factors from sparse and noisy data.~We stress that \emph{model-independent} refers to the reconstruction of the form factors:~no functional ansatz is imposed on their $t$-dependence, and the result does not rely on any particular dynamical model of hadron structure.~The word \emph{model} in \emph{diffusion model}, by contrast, refers solely to the neural-network architecture that performs the generation.~The method treats the form factor as a function discretized on a dense grid and casts the reconstruction as a conditional generation problem, analogous to image inpainting with only a small fraction of observed pixels.~The central idea is to construct the training prior from a large and diverse ensemble of synthetic form-factor curves drawn from multiple distinct functional families, each grounded in a different theoretical or phenomenological approach to hadron structure.~The resulting framework is general and can, in principle, be applied to any form factor for which such a physics-motivated prior can be assembled.~As a first and stringent test, we apply it to the three gravitational form factors of the proton, $A(t)$, $J(t)$, and $D(t)$, conditioning on lattice QCD data from Ref.~\cite{Hackett:2023nkr} and, where appropriate, on constraints from the Poincar\'{e} algebra and ChPT~\cite{Alharazin:2020yjv,Alharazin:2022wjj}.

Our main results are as follows.~First, the diffusion model provides a fully non-parametric reconstruction of all three GFFs at the unphysical pion mass of the lattice ensemble, consistent with the lattice data and with standard parametric fits across the full kinematic range $0\le -t\le 2~\mathrm{GeV}^{2}$.~A systematic data-ablation protocol demonstrates that the reconstruction remains stable even when only one or two conditioning points are retained, revealing the strong constraining power of the physically informed training prior.~Second, the continuous, covariance-resolved output of the diffusion model enables a direct extraction of the leading unknown low-energy constants of the chiral Lagrangian, $c_8$ and~$c_9$, by matching to chiral perturbation theory in the low-$|t|$ region.~The resulting values are in agreement with the model-independent dispersive determination of Cao et al.~\cite{Cao:2025dkv}, despite the two approaches being methodologically entirely distinct.~Third, using the extracted LECs to evaluate the chiral expressions at the physical pion mass, we obtain a prediction for the nucleon $D$-term, $D(0)=-4.3\pm 0.8$, consistent with the ChPT bound of Ref.~\cite{Gegelia:2021wnj} and with existing lattice and dispersive extractions.

The remainder of this paper is organised as follows.~In Sec.~\ref{sec:background} we provide a brief background on the gravitational form factors of hadrons.~In Sec.~\ref{sec:classes} we describe the construction of the training dataset from ten classes of physically motivated functional forms and their convex combinations.~Section~\ref{sec:diffusion} presents the DDPM architecture, including the $v$-prediction formulation, the dual conditioning mechanism, and the error-propagation scheme.~In Sec.~\ref{sec:reconstruction} we demonstrate the model-independent reconstruction of the proton GFFs at unphysical pion mass and extract the LECs $c_8$ and~$c_9$.~In Sec.~\ref{sec:GFFs} we reconstruct the GFFs at physical pion mass.~Network architecture, training procedure, ablation studies on synthetic test curves are collected in Appendices \ref{app:network} and \ref{app:training}.

\section{Gravitational form factors of hadrons}
\label{sec:background}
The matrix element of the conserved and symmetric EMT, $T^{\mu \nu}$, between hadronic states of the same species can be parametrized in terms of independent, conserved, Lorentz-invariant structures.~The corresponding coefficients are the GFFs.~A comprehensive and systematic analysis of the parametrization of EMT matrix elements for arbitrary spin in terms of GFFs is provided in Ref.~\cite{Cotogno:2019vjb}.

As mentioned in the introduction, the GFFs are of broad theoretical and phenomenological interest.~Their Fourier transforms admit an interpretation in terms of spatial density distributions inside hadrons.~Moreover, they are directly connected to GPDs and are, in low-energy regimes, accessible within ChPT when applicable to the hadronic system under consideration.~These features enable applications that extend beyond hadronic physics, including astrophysical phenomena such as neutron star mergers~\cite{LIGOScientific:2017vwq}, connections to fundamental aspects of General Relativity (see, e.g., Ref.~\cite{Avelino:2019esh}), and applications to the theory of hard exclusive processes and physics of exotic hadro-charmonia \cite{Dubynskiy:2008mq,Eides:2015dtr,Perevalova:2016dln}.

To illustrate some of these features we consider the matrix element of the EMT between spin-$\tfrac{1}{2}$ one-particle states, such as the nucleon, which can be decomposed as \cite{Kobzarev:1962wt,Pagels:1966zza,Ji:1996ek}:
\begin{equation}
\label{eq:emt-decomposition}
\langle p' | T^{\mu\nu}(0) | p \rangle 
= \bar{u}(p') \!\left[
  A(t)\,\frac{P^\mu P^\nu}{m_N}
+ J(t)\,\frac{i\,(P^\mu \sigma^{\nu\alpha} + P^\nu \sigma^{\mu\alpha})\Delta_\alpha}{2\,m_N}
+ D(t)\,\frac{\Delta^\mu\Delta^\nu - g^{\mu\nu}\Delta^2}{4\,m_N}
\right]\! u(p)\,,
\end{equation}
where $P = (p+p')/2$ is the average momentum, $\Delta = p'-p$ is the momentum transfer with $t = \Delta^2 < 0$ in the spacelike region, $m_N$ is the nucleon mass, and $\sigma^{\mu\nu} = \tfrac{i}{2}[\gamma^\mu,\gamma^\nu]$.~The three form factors $A(t)$, $J(t)$, and $D(t)$ are scalar functions of the Lorentz invariant~$t$ and carry direct physical meaning:

\begin{itemize}[leftmargin=0pt]
\item $A(t)$ is related to the distribution of energy and momentum inside the nucleon.~Conservation of the total four-momentum of the nucleon requires $A(0)=1$.

\item $J(t)$ encodes the distribution of total angular momentum.~For a spin-$\tfrac{1}{2}$ state, the Poincar\'e algebra fixes $J(0)=\tfrac{1}{2}$.

\item $D(t)$ governs the spatial distribution of internal forces acting inside the hadron~\cite{Polyakov:2002yz,Polyakov:2018zvc}.~Its forward-limit value $D(0)$, the so-called D-term, is not constrained by any spacetime symmetry and must be determined dynamically.~It is considered a fundamental mechanical property of the nucleon, on par with its mass and charge.
\end{itemize}
Various approaches and choices of coordinate frames exist for defining spatial density distributions from these GFFs; see, e.g., Refs.~\cite{Jaffe:2020ebz,Panteleeva:2022uii,Polyakov:2018zvc,Ernst:1960zza,Sachs:1962zzc,Miller:2018ybm,Miller:2007uy,Lorce:2020onh,Epelbaum:2022fjc,Freese:2022fat,Panteleeva:2022khw,Lorce:2018egm}.

The above GFFs are not directly measurable in scattering off gravitons but can be extracted indirectly through their connection to GPDs via the second Mellin moments~\cite{Muller:1994ses, Ji:1996ek,Radyushkin:1996nd} as follows:
\begin{eqnarray}
\int_{-1}^1 dx\,x\, H(x, \xi, t) = A(t) + \xi^2 D(t)\,, \quad \int_{-1}^1 dx\,x\, E(x, \xi, t) = 2\,J(t) - A(t) - \xi^2 D(t)\,,
\end{eqnarray}
where $H$ and $E$ are the unpolarized GPDs, $x$ is the average longitudinal momentum fraction, and $\xi$ is the skewness variable.~Determining the GFFs, and in particular the D-term, from either experiment or first-principles calculations remains challenging:~current experimental extractions, whether from DVCS~\cite{Burkert:2021ith} or from elastic scattering data (see e.g., \cite{Hashamipour:2022noy}), rely on specific GPD parametrizations, while lattice QCD computations yield results at discrete values of $t$ with sizable uncertainties near $t=0$.~Reconstructing the full $t$-dependence from such sparse and noisy data without committing to a specific functional ansatz is the central goal of the present work.~To this end, we construct a diverse training prior from multiple families of physically motivated functions, described in the next section.

\section{Training data: classes of functional forms}
\label{sec:classes}

The diffusion model learns the manifold of physically plausible GFF curves from a training set of $6\times10^{5}$ synthetic form factors, sampled on a uniform grid of $N_t=200$ points in $0\le -t\le 2~\mathrm{GeV}^{2}$.
The training curves are drawn from ten classes of functional forms: eight rooted in distinct theoretical or phenomenological approaches to hadron form factors, and two built from convex combinations that populate the gaps in shape space between the theory-motivated families.
For each class, the free parameters are sampled from multiple randomized strategies, typically five to nine per parameter, covering uniform, log-uniform, Gaussian, physically motivated, and correlated joint distributions, ensuring dense coverage of the accessible shape space without overpopulating any single region.
The forward-limit value ${\rm GF}(0)$ is drawn from a broad mixture of distributions spanning $[-10,\,10]$.
All curves are subject to the acceptance criteria $|{\rm GF}(t)|\le 10$, absence of numerical singularities, and a minimum variation threshold that removes effectively constant functions.

The eight base functional forms are:
\begin{subequations}
\label{eq:families}
\begin{align}
\text{Multipole:}~~
{\rm GF}(t) &= \frac{{\rm GF}(0)}{\bigl(1 - t/M^2\bigr)^{n}}\,,
&\quad
\text{$z$-exp.:}~~
{\rm GF}(t) &= \sum_{k=0}^{K} a_k\, z(t)^{k}\,,
\label{eq:multipole_zexp}\\[4pt]
\text{Meson dom.:}~~
{\rm GF}(t) &= \sum_{i=1}^{N} \frac{c_i}{m_i^{2} - t}\,,
&\quad
\text{Mod.\ exp.:}~~
{\rm GF}(t) &= {\rm GF}(0)\,(1\! -\! t/\Lambda^{2})^{\beta}\,
       e^{\gamma t/\Lambda^{2}}\,,
\label{eq:mesdom_modexp}\\[4pt]
\text{Pad\'{e}:}~~
{\rm GF}(t) &= \frac{a_{0} + \sum\limits_{i=1}^{N} a_{i}(-t)^{i}}
            {1 + \sum\limits_{j=1}^{M} |b_{j}|(-t)^{j}}\,,
&\quad
\text{Dispersive:}~~
{\rm GF}(t) &= \frac{1}{\pi}\!\int\limits_{t_{\rm cut}}^{s_{\rm max}} \!\!ds\,
       \frac{\rho(s)}{s\! -\! t}\,,
\label{eq:pade_disp}\\[4pt]
\text{Log-mod.:}~~
{\rm GF}(t) &= \frac{{\rm GF}(0)}{(1\! -\! t/M^{2})^{n}}\,
       [1\! +\! c\ln(1\! -\! t/M^{2})]^{\delta}\!,
&\quad
\text{Bag:}~~
{\rm GF}(t) &= {\rm GF}(0)\,\frac{3j_{1}(R\sqrt{-t})}
                     {R\sqrt{-t}}\,
       e^{-\beta(-t)}\,.
\label{eq:logmod_bag}
\end{align}
\end{subequations}
In Eq.~\eqref{eq:multipole_zexp}, $z(t) = (\sqrt{t_{\rm cut}-t}-\sqrt{t_{\rm cut}-t_0})/(\sqrt{t_{\rm cut}-t}+\sqrt{t_{\rm cut}-t_0})$ is the conformal variable that maps the cut $t$-plane onto the unit disk.
In Eq.~\eqref{eq:pade_disp}, the spectral function is modeled as $\rho(s) = (s-t_{\rm cut})^{p}\,\sum_{i} c_i\Gamma_i/[(s-m_i^{2})^{2}+(\Gamma_i/2)^{2}]\,e^{-s/s_{\rm max}}$, combining a threshold factor $(s-t_{\rm cut})^{p}$, Breit--Wigner peaks $\Gamma_i$, and a high-energy damping $s_{\rm max}$.

The motivation for including each class is as follows.
The \emph{multipole}~\eqref{eq:multipole_zexp} is the standard parametrization in lattice QCD extractions of GFFs~\cite{Hackett:2023nkr,Pefkou:2021fni,Alexandrou:2019ali,Bali:2018zgl} and therefore must be well-represented in the prior.
The \emph{$z$-expansion}~\eqref{eq:multipole_zexp} is the most general parametrization consistent with the analyticity of the form factor~\cite{Hill:2010yb,Bhattacharya:2011ah,Lee:2015jqa} and is used alongside the multipole in the lattice analysis of Ref.~\cite{Hackett:2023nkr}; including it ensures that the prior covers the full space of analytic functions, not only the pole-dominated subset.
Because $|z_{\rm max}|\approx 0.39$ on the physical grid, naive coefficient sampling produces nearly linear curves; we compensate it by rescaling $a_k\to a_k/(|z_{\rm max}|^{k})^\alpha$, $\alpha \in [0.3, 1]$, so that each order contributes $\mathcal{O}(1)$ to the shape variation.
\emph{Meson dominance}~\eqref{eq:mesdom_modexp} represents the GFF as a sum of $t$-channel meson exchanges with the appropriate quantum numbers \cite{Broniowski:2024oyk,Masjuan:2012sk,Bijker:2004yu}; for ${\approx}\,70\%$ of the $N\ge 3$ samples the residues satisfy the superconvergence sum rule $\sum_i c_i = 0$~\cite{Broniowski:2024oyk,Tong:2022zax}, guaranteeing the correct pQCD falloff ${\rm GF}(t)\sim 1/(-t)^{2}$.
The \emph{modified exponential}~\eqref{eq:mesdom_modexp} captures the Gaussian-like shapes that arise in constituent-quark models and soft-wall AdS/QCD~\cite{Brodsky:2008pf,Abidin:2009hr}, with the polynomial prefactor $(1-t/\Lambda^{2})^{\beta}$ resolving the vanishing-slope problem of the pure exponential at $t=0$.
\emph{Pad\'{e} approximants}~\eqref{eq:pade_disp} provide a systematic rational-function generalization that encompasses multipoles and the Kelly parametrization~\cite{Kelly:2004hm,Ye:2017gyb,Schlessinger:1966zz,Cui:2021vgm}; the use of $|b_j|$ in the denominator guarantees the absence of spacelike poles by construction, and the condition $M\ge N+2$ (imposed for ${\approx}\,75\%$ of samples) enforces the correct large-$(-t)$ asymptotics.
The \emph{dispersive representation}~\eqref{eq:pade_disp} exploits analyticity through an unsubtracted dispersion relation, following the approach of Refs.~\cite{Cao:2025ncp,Hammer:2003ak,Hohler:1976ax,Belushkin:2006qa}; its inclusion ensures that the prior contains curves shaped by the two-pion threshold and resonance interference patterns that are absent from the purely algebraic families.
The \emph{log-modified multipole}~\eqref{eq:logmod_bag} incorporates the logarithmic corrections from the running of $\alpha_s$ at large $-t$~\cite{Tong:2022zax,Brodsky:1973kr,Lepage:1980fj}, recovering the pure multipole in the limit $c\to 0$ or $\delta\to 0$.
Finally, the \emph{bag-model / Bessel-type} form~\eqref{eq:logmod_bag} is motivated by the MIT bag model~\cite{Chodos:1974je,Ji:1997gm,Neubelt:2019sou,Tezgin:2024tfh}, whose GFFs arise from Fourier transforms of step-function-like spatial distributions; its characteristic mild shoulder near $-t\sim 1/R^{2}$ introduces a qualitative feature absent from all smooth monotone families.

To fill the remaining gaps in shape space between the eight base classes, we generate two additional families of convex combinations.~Class~9 consists of pairwise combinations $\lambda\,f_A + (1-\lambda)\,f_B$ with $\lambda\sim\mathrm{U}(0,1)$, drawn from all $\binom{8}{2}=28$ pairs.~Class~10 uses Dirichlet-weighted mixtures of $M$ randomly chosen parent families, with the concentration parameter itself randomized to interpolate between near-uniform and peaked weightings.~In total, $6\times10^{4}$ accepted curves are generated per class, yielding $6\times10^{5}$ training samples; a separate validation set of $6\times10^{3}$ curves is held out for model selection.~Each curve is normalized to zero mean and unit variance per grid point before training.

The complete generation code, including all parameter sampling strategies, acceptance criteria and analysis of the final results, is provided in Ref.~\cite{GitHub} in fully reproducible form.
\section{Denoising Diffusion Probabilistic Model}
\label{sec:diffusion}

We employ a denoising diffusion probabilistic model
(DDPM)~\cite{Ho:2020ddpm} with $v$-prediction~\cite{Salimans:2022whc}
to reconstruct GFF curves from sparse data.
This section summarises the formalism, for a comprehensive treatment
see Refs.~\cite{Ho:2020ddpm,Nichol:2021improved,Salimans:2022whc}.

\subsection{Forward process and \texorpdfstring{$v$}{v}-prediction}
\label{sec:diff:forward}

Let $x_0\in\mathbb{R}^{L}$ ($L=200$) denote a GFF curve sampled on a uniform grid.~The forward (noising) process produces a noisy version at diffusion step $\tau\in\{1,\dots,T\}$, with $T{=}1000$, via
\be
\label{eq:forward}
  x_\tau = \sqrt{\bar\alpha_\tau}\;x_0
         + \sqrt{1-\bar\alpha_\tau}\;\varepsilon\,,
  \qquad \varepsilon\sim\mathcal{N}(0,\mathbf{I})\,, \quad \bar\alpha_\tau=\prod_{s=1}^{\tau}\alpha_s\,, \quad \alpha_s=1-\beta_s\,,
\ee
where $\varepsilon$ denotes the pure noise drawn from a standard normal distribution $\mathcal{N}(0,\mathbf{I})$, $\bar\alpha_\tau$ is the signal preservation coefficient and  $\beta_s\in (0,1)$ is the noise variance at step~$s$.~We adopt the cosine noise schedule of Ref.~\cite{Nichol:2021improved},
\be
\label{eq:cosine}
  \bar\alpha_\tau = \frac{f(\tau)}{f(0)}\,,
  \qquad
  f(\tau) = \cos^2\!\biggl(\frac{\tau/T+s}{1+s}\,\frac{\pi}{2}\biggr),
\ee
with offset $s=0.008$, which ensures a smooth transition from
$\bar\alpha_0\approx 1$ (pure signal) to $\bar\alpha_T\approx 0$
(pure noise).~Rather than predicting the noise $\varepsilon$ or the clean data $x_0$ directly, our network predicts the \emph{velocity}
\be
\label{eq:vtarget}
  v_\tau = \sqrt{\bar\alpha_\tau}\;\varepsilon
         - \sqrt{1-\bar\alpha_\tau}\;x_0\,,
\ee
which has uniform variance across all noise levels, avoiding the signal-to-noise imbalance that afflicts $\varepsilon$- and $x_0$-prediction at opposite ends of the diffusion schedule.~From a predicted $\hat v_\tau$ one obtains the predicted 
\bea
\label{eq:recover}
  \hat x_0 &=& \sqrt{\bar\alpha_\tau}\;x_\tau
             - \sqrt{1-\bar\alpha_\tau}\;\hat v_\tau\,,
  \\[2pt]
  \hat\varepsilon &=& \sqrt{1-\bar\alpha_\tau}\;x_\tau
                    + \sqrt{\bar\alpha_\tau}\;\hat v_\tau\,,
\eea
which can be verified by substituting Eqs.~(\ref{eq:forward})--(\ref{eq:vtarget}).~The relations in Eq.~(\ref{eq:recover}) are exact identities:~given the true $v_\tau $, one recovers $x_0$ and $\varepsilon$ without approximation.~At inference time the network provides an approximation $\hat v_\tau$, so the recovered quantities are likewise approximated.

\subsection{Reverse sampling}
\label{sec:diff:reverse}
To generate a GFF curve one must run the noising process in reverse, stepping from pure noise $x_T$ back to clean data $x_0$.~The one-step reverse distribution $q(x_{\tau-1}\mid x_\tau)$, the probability of the less noisy state given the current noisy state, is intractable, since it depends on the unknown data distribution.~However, if one additionally conditions on the clean data $x_0$, Bayes' theorem yields a closed-form Gaussian \mbox{(see, e.g., Ref.~\cite{Ho:2020ddpm})}:
\be
\label{eq:posterior}
  q(x_{\tau-1}\mid x_\tau,\,x_0)
  = \mathcal{N}\!\bigl(x_{\tau-1};\;
    \tilde\mu_\tau,\;\tilde\beta_\tau\,\mathbf{I}\bigr)\,,
\ee
referred to as the \emph{posterior}\,\footnote{In the Bayesian sense:~$q$ is the distribution of $x_{\tau-1}$ \emph{after} conditioning on the observed quantities $x_\tau$ and $x_0$.~This is standard terminology in the generative-modelling literature~\cite{Ho:2020ddpm,Nichol:2021improved,Salimans:2022whc}.},
with mean and variance
\bea
\label{eq:posterior_mean}
  \tilde\mu_\tau
  &=& \frac{\sqrt{\bar\alpha_{\tau-1}}\,\beta_\tau}
           {1-\bar\alpha_\tau}\;x_0
    + \frac{\sqrt{\alpha_\tau}\,(1-\bar\alpha_{\tau-1})}
           {1-\bar\alpha_\tau}\;x_\tau\,,
  \\[4pt]
\label{eq:posterior_var}
  \tilde\beta_\tau
  &=& \frac{\beta_\tau\,(1-\bar\alpha_{\tau-1})}
           {1-\bar\alpha_\tau}\,.
\eea

Of course $x_0$ is unknown at inference time, it is precisely what we wish to reconstruct.~The neural network $f_\theta$ fills this gap: at each step it predicts $\hat v_\tau$, from which an estimate $\hat x_0$ is obtained via Eq.~(\ref{eq:recover}) and substituted for $x_0$ in Eq.~(\ref{eq:posterior_mean}).~Starting from $x_T\sim\mathcal{N}(0,\mathbf{I})$, each reverse step reads\footnote{The fresh noise $z$ injected at each step is the source of sample diversity:~different random seeds yield different plausible GFF curves whose spread constitutes the model's uncertainty band.}
\be
\label{eq:reverse_step}
  x_{\tau-1} = \tilde\mu_\tau + \sqrt{\tilde\beta_\tau}\;z\,,
  \qquad z\sim\mathcal{N}(0,\mathbf{I})\,,
  \qquad \tau=T,\dots,2\,,
\ee
with $x_0=\tilde\mu_1$ at the final step (no noise added).

\subsection{Conditioning and error propagation}
\label{sec:diff:conditioning}
Our goal is to reconstruct the full GFF  curve  given $K$ known data points\footnote{Here $F_k$ refers to a (gravitational) form factor $F(t_k)$.} $\{(t_k,\,F_k\pm\sigma_k)\}_{k=1}^{K}$ from, e.g., lattice QCD or ChPT.~The model conditions on this information through two complementary mechanisms.~First, we have the concatenation conditioning:~each known value is mapped to the nearest grid index~$j_k$ and normalised to the model's internal representation:~$\tilde F_k = (F_k - \mu_{j_k})/\sigma_{j_k}$, where $\mu$ and $\sigma$ are the per-grid-point mean and standard deviation of the training set.~The network receives three input channels at every reverse step:~the current noisy estimate $x_\tau$, a binary mask $m\in\{0,1\}^L$ with $m_{j_k}=1$ at known positions, and a condition vector $c\in\mathbb{R}^L$ with $c_{j_k}=\tilde F_k$ and zero elsewhere.~During training, random masks (50\% uniformly random, 30\% clustered in the low-$|t|$ region, 20\% unconditional) teach the network to reconstruct missing values conditioned on an arbitrary subset of known points.~To propagate the uncertainties of the known data into the reconstructed uncertainty band, each of the $N$ generated samples receives its own realisation of the conditioning values\footnote{It is worth mentioning that throughout the lattice reconstructions presented in the next section, the uncertainties of the conditioning points are propagated independently  via
Eq.~(\ref{eq:jitter}), as the full covariance matrix of 
Ref.~\cite{Hackett:2023nkr} is publicly not available;~the impact of 
off-diagonal elements is expected to be subdominant, since the 
reconstruction remains stable under data reduction, as we will see below.},
\be
\label{eq:jitter}
  c_n^{(j_k)} = \tilde F_k + \xi_n\,,
  \qquad
  \xi_n \sim \mathcal{N}\!\bigl(0,\,\tilde\sigma_k^{\,2}\bigr)\,,
  \qquad
  \tilde\sigma_k = \sigma_k / \sigma_{j_k}\,,
\ee
where $\sigma_k$ is the error of the known data and $\tilde\sigma_k$ its normalised counterpart.~This noise injection is drawn once per sample and held fixed throughout the reverse chain: both the concatenation input~$c$ and the replacement values use the same $c_n$.~Then we have the replacement part.~At each intermediate reverse step $\tau\to\tau{-}1$, the known grid points in $x_{\tau-1}$ are overwritten by a forward-noised version of the conditioning values at the appropriate noise level:
\be
\label{eq:replacement}
  x_{\tau-1}^{(j)} \;\longleftarrow\;
  \begin{cases}
    \sqrt{\bar\alpha_{\tau-1}}\;c_n^{(j)}
    + \sqrt{1-\bar\alpha_{\tau-1}}\;\eta_\tau\,,
    & j\in\{j_k\}\,,\\[2pt]
    x_{\tau-1}^{(j)}\,,
    & \text{otherwise}\,,
  \end{cases}
\ee
where $\eta_\tau\sim\mathcal{N}(0,\mathbf{I})$ is drawn independently at each reverse step.~At the final step ($\tau=1$) the clean conditioning values are inserted directly: $x_0^{(j_k)}\leftarrow c_n^{(j_k)}$.

    It is important to emphasise that the two conditioning channels, concatenation and replacement, play complementary roles.~The concatenation input $(m,c_n)$ informs the network \emph{globally} at every denoising step, allowing it to shape the predicted $\hat v_\tau$ over the entire grid in light of the known data.~The replacement step enforces \emph{local} consistency by anchoring the known grid points to their correct marginal distribution $q(x_{\tau-1}^{(j_k)}\mid c_n^{(j_k)})$ at each noise level, eliminating any residual drift at those positions.~Because every sample~$n$ uses a different noised realisation $c_n$, the empirical spread of the $N$ generated curves encodes both the model's learned prior uncertainty and the propagated experimental errors.

Our network $f_\theta$ was trained to minimise the mean-squared error on the $v$-target,
\be
\label{eq:loss}
\mathcal{L}(\theta) = \mathbb{E}_{x_0,\,\varepsilon,\,\tau,\,m}\! \bigl[\lVert f_\theta(x_\tau,\tau,m,c) - v_\tau \rVert^2\bigr]\,.
\ee
Further details about the network architecture and training are given in Appendices \ref{app:network} and \ref{app:training}.~The trained model, sampling scripts, and all configuration details required to reproduce the results presented in this work are publicly available in Ref.~\cite{GitHub}.

For each reconstruction presented below, $N=10^4$ curves are generated from the reverse process.~The pointwise 68\% credible interval is defined by the 16th and 84th percentiles of the ensemble at each grid point.~A jackknife resampling of the ensemble mean yields standard errors that are indistinguishable from the median curve at the scale of all figures shown; we therefore report the median as the central estimate throughout.

\section{Model-independent reconstruction of the proton GFFs at unphysical pion mass}
\label{sec:reconstruction}
In this section, we demonstrate that the diffusion-based generative framework yields a fully \emph{model-independent} reconstruction of the total proton GFFs $A(t)$, $J(t)$, and $D(t)$, consistent with the lattice QCD results of Ref.~\cite{Hackett:2023nkr} across the full kinematic range $0 \le -t \le 2~\mathrm{GeV}^{2}$, without imposing any functional form on the $t$-dependence.~The only inputs to the reconstruction are:

\begin{enumerate}
\item A limited small number of lattice data points from the total GFF results of Ref.~\cite{Hackett:2023nkr}.
\item Exact physical constraints that follow from the symmetry structure of the EMT, or, where no such constraint exists, well-motivated sign conditions supported by theoretical arguments and the data themselves.
\end{enumerate}
Having learned the space of physically admissible form-factor shapes 
from the training prior (Sec.~\ref{sec:classes}), the diffusion model samples from the posterior distribution consistent with the supplied conditioning input, yielding a non-parametric estimate of both the central value and the pointwise uncertainty of the reconstructed form factor.

To probe the robustness of the reconstruction, we adopt a systematic \emph{data-ablation protocol}:~starting from the most informative conditioning set, we progressively remove lattice data points until only a single point and the relevant physical constraint remain (for $D(t)$ we retain two points).~This protocol reveals how much information each data point contributes, exposes the regions of $t$ where the reconstruction is driven by the physical prior rather than the data, and provides
a transparent, data-driven quantification of the extrapolation uncertainty that parametric fits absorb into the assumed functional form.

We emphasize that for each available lattice dataset we select as conditioning input those points with only modest statistical uncertainties compared with the others, in other words, the reconstruction is not aided by large error bars that would trivially accommodate many functional shapes.~The diffusion model must reproduce sharply defined values around the conditioning points while simultaneously generating physically sensible interpolations and extrapolations throughout the kinematic range.~That it succeeds with as few as one or two data points is a direct consequence of the physically informed prior built into the training data, and constitutes a stringent validation of the approach.

\subsection{Reconstruction of symmetry-constrained form factors~$A(t)$ and $J(t)$}
\label{sec:AJ_reconstruction}
\begin{figure}[t]
  \centering
  \begin{subfigure}[t]{0.48\textwidth}
    \centering
    \includegraphics[width=\textwidth]{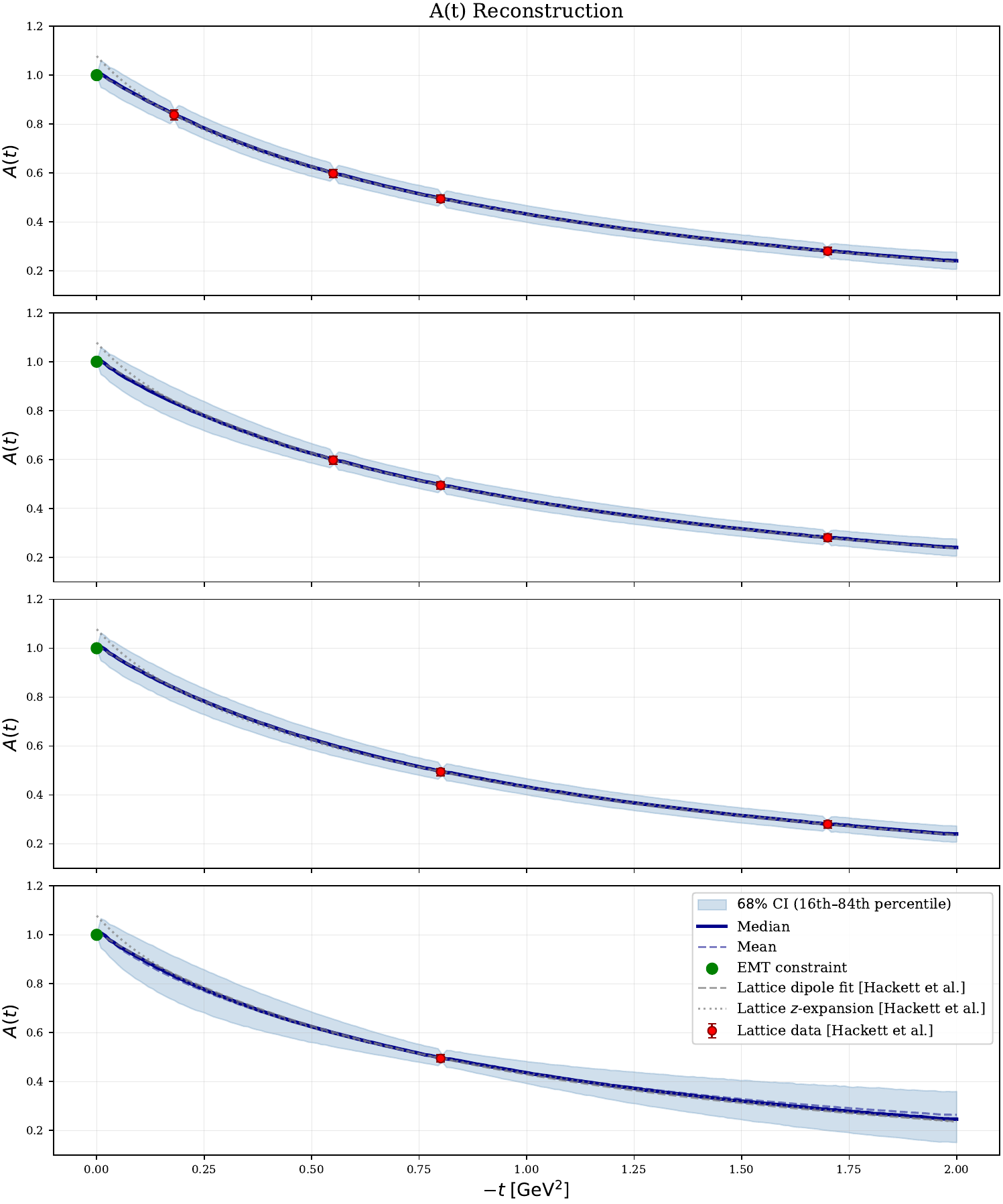}
    \caption{}
    \label{fig:A_reconstruction}
  \end{subfigure}
  \hfill
  \begin{subfigure}[t]{0.48\textwidth}
    \centering
    \includegraphics[width=\textwidth]{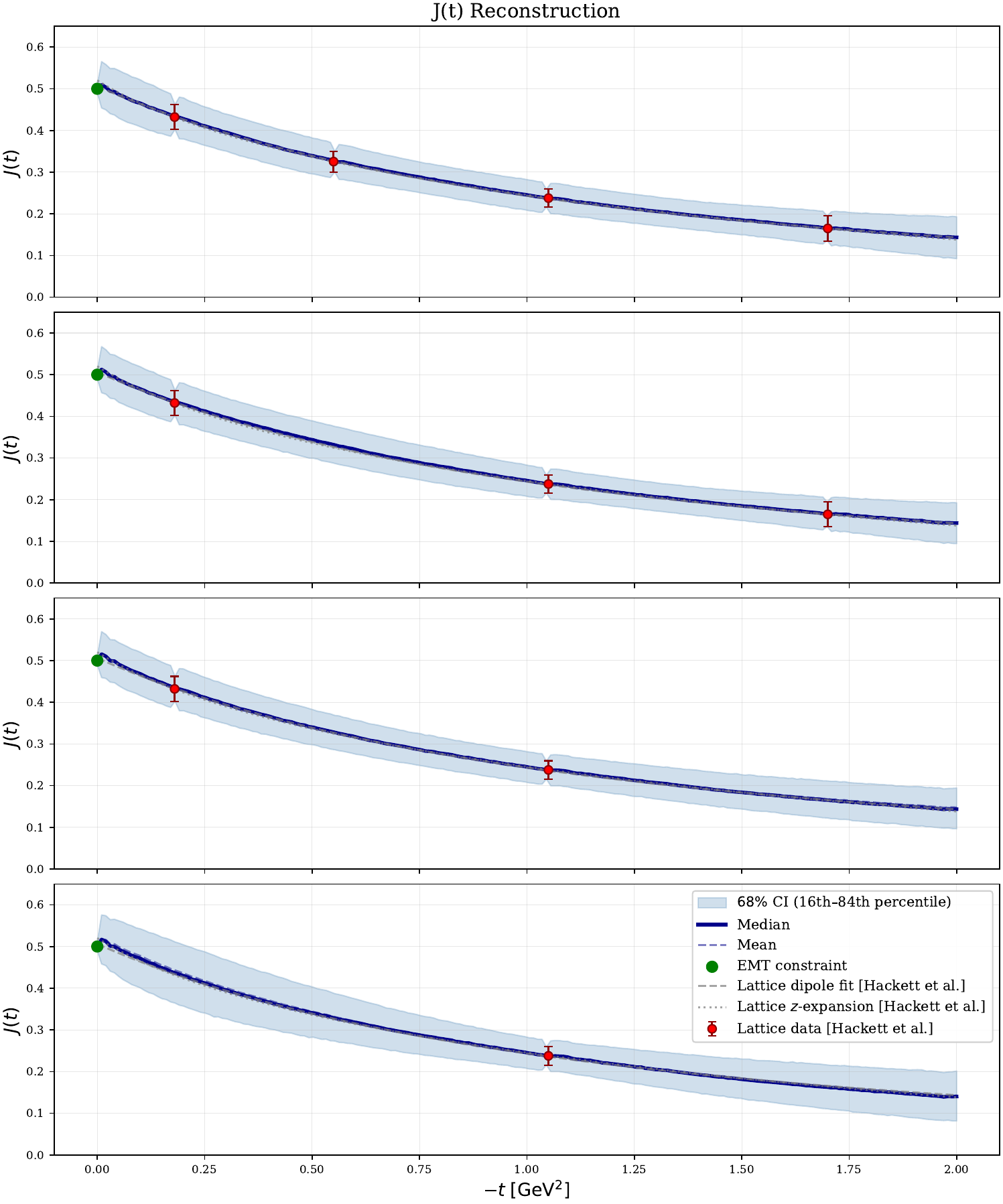}
    \caption{}
    \label{fig:J_reconstruction}
  \end{subfigure}
  \caption{Model-independent reconstruction of $A(t)$ and $J(t)$.~In each column, the four panels correspond to progressively sparser conditioning:~the top panel uses the largest subset of lattice data points (red markers), and successive panels remove points until only a single datum remains.~The green marker denotes the constraint imposed by the Poincar\'{e} algebra.~Gray dashed and dotted curves are the dipole and $z$-expansion fits of Ref.~\cite{Hackett:2023nkr}.}
\label{fig:AJ_reconstruction}
\end{figure}

The form factors $A(t)$ and $J(t)$ benefit from exact forward-limit constraints that follow from the Poincar\'{e} algebra:
\begin{equation}
A(0) = 1\,,\qquad J(0) = \tfrac{1}{2}\,.
\label{eq:sumrules}
\end{equation}
The first one expresses conservation of total four-momentum, and $J(0)=1/2$ follows from the proton being a spin-$\tfrac{1}{2}$ state whose total angular momentum is encoded in the forward matrix element of the EMT.~That is, we impose them as exact conditioning points alongside a small number of lattice data points.~The results of the data-ablation study are presented in Fig.~(\ref{fig:AJ_reconstruction}), we observe that:
\begin{itemize}[leftmargin=0pt]
\item \textit{Full conditioning}:~When conditioned on 4 lattice data points plus the forward-limit constraint (top panels), the diffusion-model median is virtually indistinguishable from both the dipole and z-expansion fits of Ref.~\cite{Hackett:2023nkr} across the entire kinematic range.~The 68\% credible band tightly envelops both parametrizations and, consequently, also encompasses all remaining lattice data points for $A(t)$ and $J(t)$ reported in Fig.~(1) of Ref.~\cite{Hackett:2023nkr} that were \emph{not} used as conditioning input, confirming that the model has captured the correct shape of both form factors without being informed of any functional form.
\item \textit{Stability under data reduction}:~As conditioning points are progressively removed (second through third panels), the median remains remarkably stable and continues to track the parametric fits.~The credible interval broadens precisely in the regions from which data have been removed, predominantly at large $-t$, providing a faithful, data-driven measure of the growing extrapolation uncertainty.
\item \textit{Reconstruction from minimal input}:~The most notable result appears in the bottom panels, where {only a single lattice data point at intermediate $-t$ and the forward-limit constraint} are supplied.~Even in this extreme scenario the reconstruction yields a smoothly decaying curve whose median remains compatible with the
full parametric fits.~At $-t = 2~\mathrm{GeV}^{2}$ the credible band remains moderate for both form factors.~The notably narrow uncertainty obtained with only two conditioning inputs, despite the wide range of mathematically allowed curves through those points, is a direct consequence of the physical prior encoded in the training data.

\end{itemize}
\subsection{Reconstruction of D form factor}
\label{sec:D_reconstruction}

\begin{figure}[t]
  \centering
  \begin{subfigure}[t]{0.48\textwidth}
    \centering
    \includegraphics[width=\textwidth]{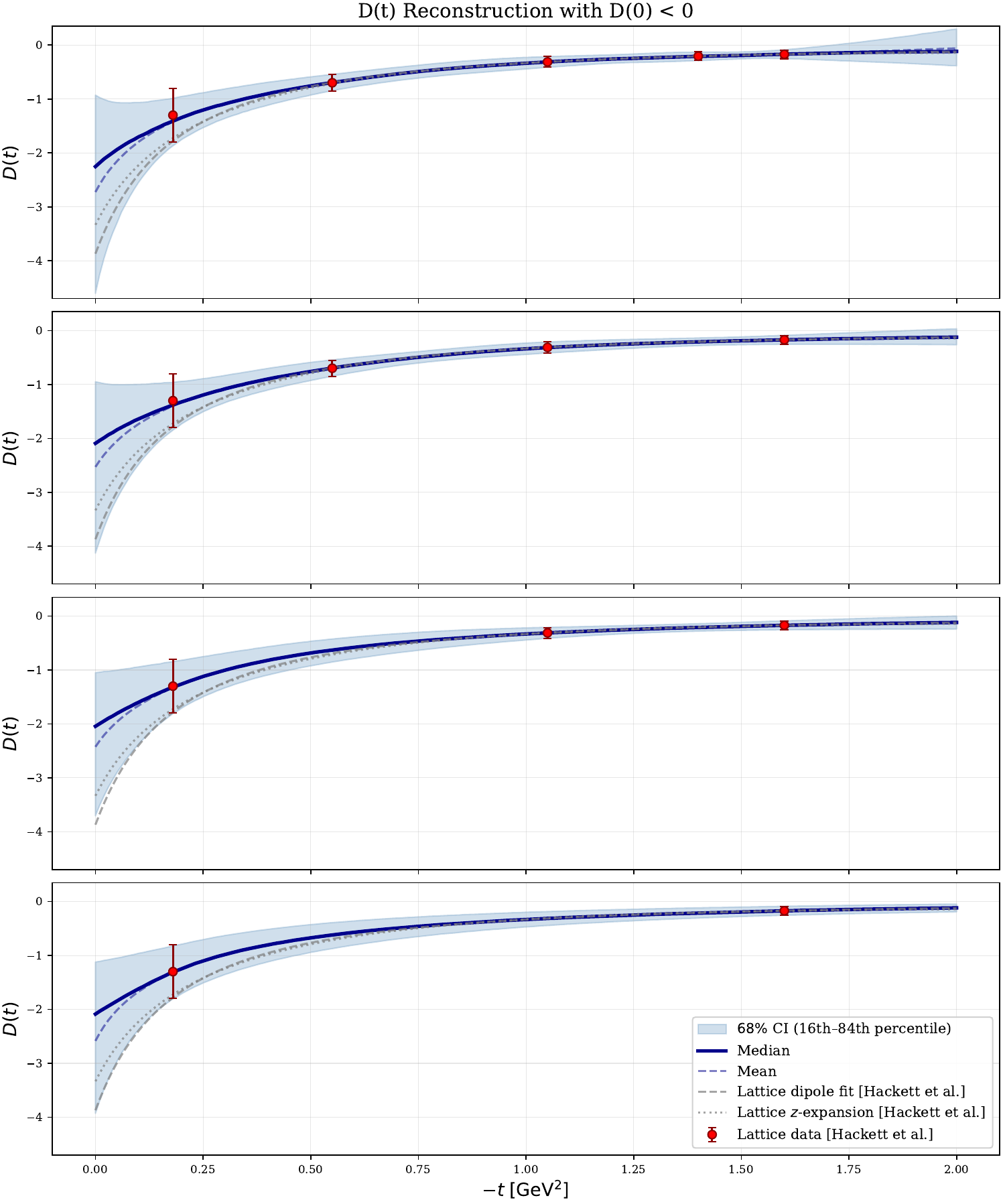}
    \caption{}
    \label{fig:D_D0less0}
  \end{subfigure}
  \hfill
  \begin{subfigure}[t]{0.48\textwidth}
    \centering
    \includegraphics[width=\textwidth]{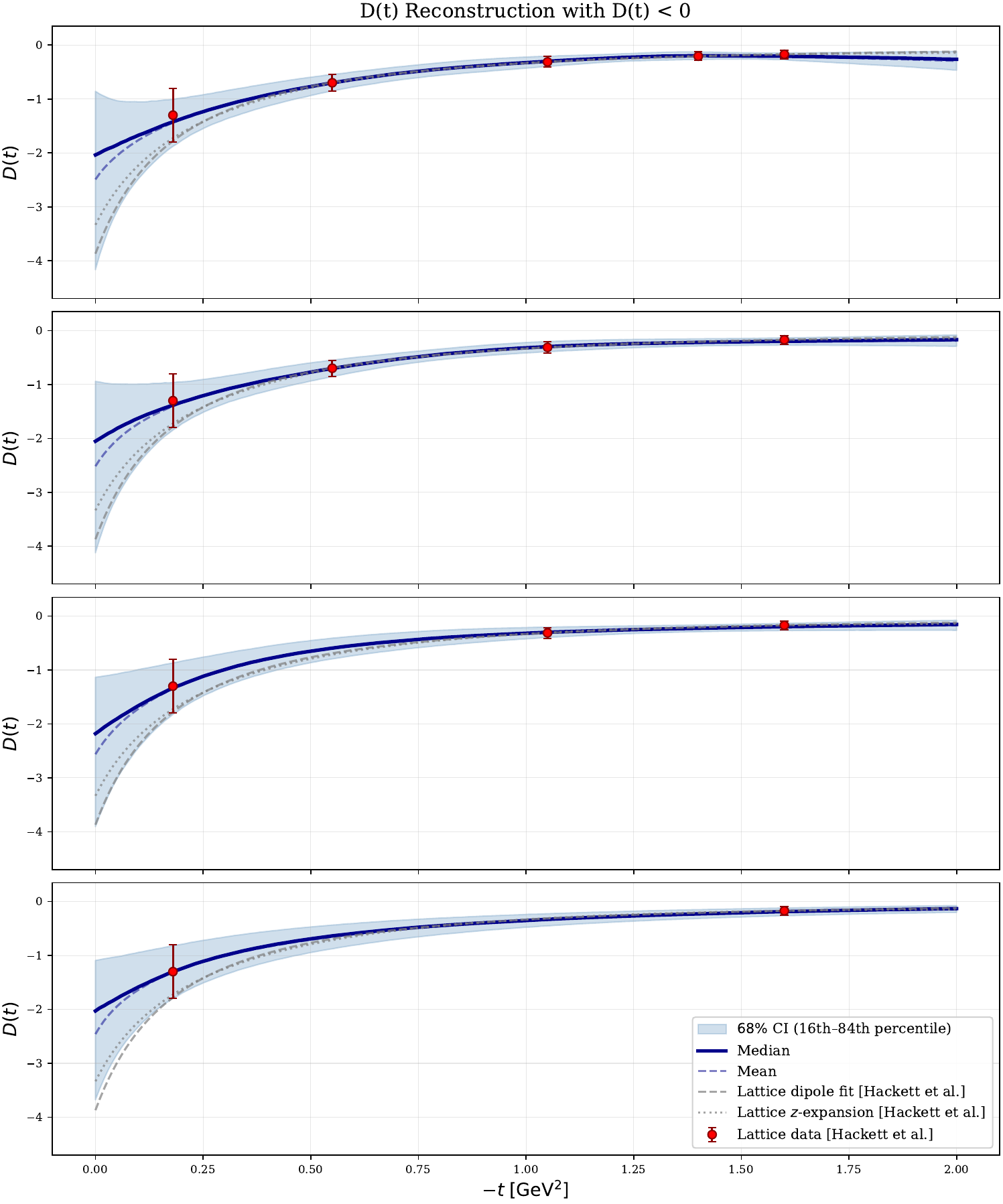}
    \caption{}
    \label{fig:D_Dtless0}
  \end{subfigure}
  \caption{Reconstruction of the $D(t)$ under two physics-informed criteria.~\textbf{(a)}~Only D-term is negative.~\textbf{(b)}~Negativity is enforced at every point on the spacelike grid:~$D(t)<0$.~Panel layout, markers, and bands are as in Fig.~(\ref{fig:AJ_reconstruction}).}
 \label{fig:D_reconstruction}
\end{figure}
The form factor $D(t)$ presents the most stringent test of the reconstruction framework, for two reasons.~First, unlike $A(t)$ and $J(t)$, the forward limit $D(0)$ is not fixed by any spacetime symmetry, i.e., it is a quantity that must be determined from experiment or first-principles calculation.~Second, in the low-energy region relevant for the extrapolation to $D(0)$, the lattice data for $D(t)$ exhibit substantially larger statistical uncertainties than those for $A(t)$ and $J(t)$.~In combination, these features lead to a strong \textit{sensitivity} of $D(0)$ to the choice of fitting ansatz.

In the absence of a sum rule, we impose physically motivated sign constraints.~From the raw diffusion-model ensemble we retain only those samples satisfying the prescribed sign condition, without modifying the generative process.~Two variants of constraints are considered:
\begin{itemize}[leftmargin=0pt]
\item Minimal constraint ($D(0) < 0$): The requirement that $D(0)$ be negative  follows from the mechanical stability of the proton~\cite{Polyakov:2018zvc,Perevalova:2016dln}:~a positive $D$-term would imply a repulsive normal force and the system would collapse.~This negativity is  also supported by ChPT~\cite{Gegelia:2021wnj}, chiral quark-soliton models~\cite{Goeke:2007fp,Cebulla:2007ei}, and the lattice data themselves.~This weakest condition allows the form factor to fluctuate mildly at large~$-t$.
\item Full negativity ($D(t) < 0$):~Negativity is enforced at every point on the spacelike grid.~This stronger condition is motivated by all known model calculations, see e.g.~Refs.~\cite{Cebulla:2007ei, Cao:2025ncp,Yao:2024ixu,Cao:2025dkv,Deng:2025azp}.
\end{itemize}

Figure~(\ref{fig:D_D0less0}) shows the reconstruction with the
condition $D(0) < 0$ only.~In the most densely conditioned scenario (top panel), in the region $-t>0.3~\mathrm{GeV}^{2}$, where the lattice data are less noisy, the median closely tracks both parametric fits and the credible band encompasses both across the entire $-t$ range.~Requiring only $D(0)<0$ restricts the reconstructed ensemble to negative values over nearly the entire kinematic range.~A small positive deviation of the upper edge of the credible band appears only in the far high-$|t|$ tail, $-t\gtrsim1.65~\mathrm{GeV}^2$, where the form factor is close to zero and the statistical spread is largest.~However, the ensemble median and mean remain strictly negative at every value of $t$ in all four conditioning scenarios.~Imposing the stronger condition $D(t)<0$ at every grid point (Fig.~(\ref{fig:D_Dtless0})) removes this residual fraction of samples and yields a modest tightening of the band for $-t\gtrsim1.65~\mathrm{GeV}^2$.~Away from this narrow corner of the kinematic range the two conditioning schemes are essentially indistinguishable.~In particular, the median, mean, and credible interval in the vicinity of $t=0$, where the physical interest in $D(0)$ resides, change only marginally between Figs.~(\ref{fig:D_D0less0}) and (\ref{fig:D_Dtless0}).
This near-equivalence is non-trivial:~it demonstrates that the diffusion-model prior, shaped by the physically motivated training data, already strongly disfavors zero crossings of $D(t)$ in the spacelike region.~Requiring $D(0) < 0$ alone is therefore, within the resolution of the present analysis, sufficient to enforce negativity across the full $t$-range.

Across all four conditioning scenarios, the width of the credible band at small $-t$ remains essentially unchanged, reflecting the presence of the lowest-$|t|$ point in every reconstruction.~This behavior has a clear physical origin:~the form factor attains its largest absolute value in the forward region and must vanish at large $|t|$, so the bulk of the dynamic range is concentrated at low momentum transfer, and a single well-placed conditioning point there exerts maximal leverage on the reconstruction.~By contrast, at large $|t|$ the form factor flattens toward zero and becomes comparatively insensitive to individual data points.~This is also evident in the final panels of Fig.~(\ref{fig:AJ_reconstruction}), where retaining only the $t=0$ constraint and a single intermediate-$|t|$ lattice point already suffices to constrain the forward behavior.

These observations are corroborated by the ablation study on synthetic test curves presented in Appendix~\ref{sec:Ablation}.~That analysis demonstrates that a reliable model-independent reconstruction requires, at minimum, two well-determined points:~one in the forward region $-t \lesssim 0.25~\mathrm{GeV}^{2}$, which anchors the form-factor normalization and slope, and a second at intermediate momentum transfer $-t \sim 0.25$--$0.5~\mathrm{GeV}^{2}$, which stabilizes the interpolation into the mid-$|t|$ range.~The inclusion of a third conditioning point at higher $|t|$ further tightens the credible band but yields a comparatively smaller marginal improvement, consistent with the flattening of the form factor in that region.~Conversely, when only high-$|t|$ data are available, the reconstruction in the physically critical forward region degrades substantially (see the bottom row of Fig.~(\ref{fig:ablation_appendix})).

These findings may also be of relevance when planning future lattice QCD calculations of the gravitational form factors.~Since the model-independent reconstruction is most sensitive to the precision of data in the low- and intermediate-$|t|$ region, investments in reducing statistical uncertainties at $-t \lesssim 0.5~\mathrm{GeV}^{2}$, for instance through increased statistics at small momentum transfer, would be particularly valuable for tightening the extrapolation to $t = 0$.~Improvements at large $|t|$, while beneficial for the overall characterization of the form factor, have a more modest impact on the quantities of primary physical interest.

\begin{figure}[H]
  \centering
  \begin{subfigure}[H]{0.48\textwidth}
    \centering
    \includegraphics[width=\textwidth]{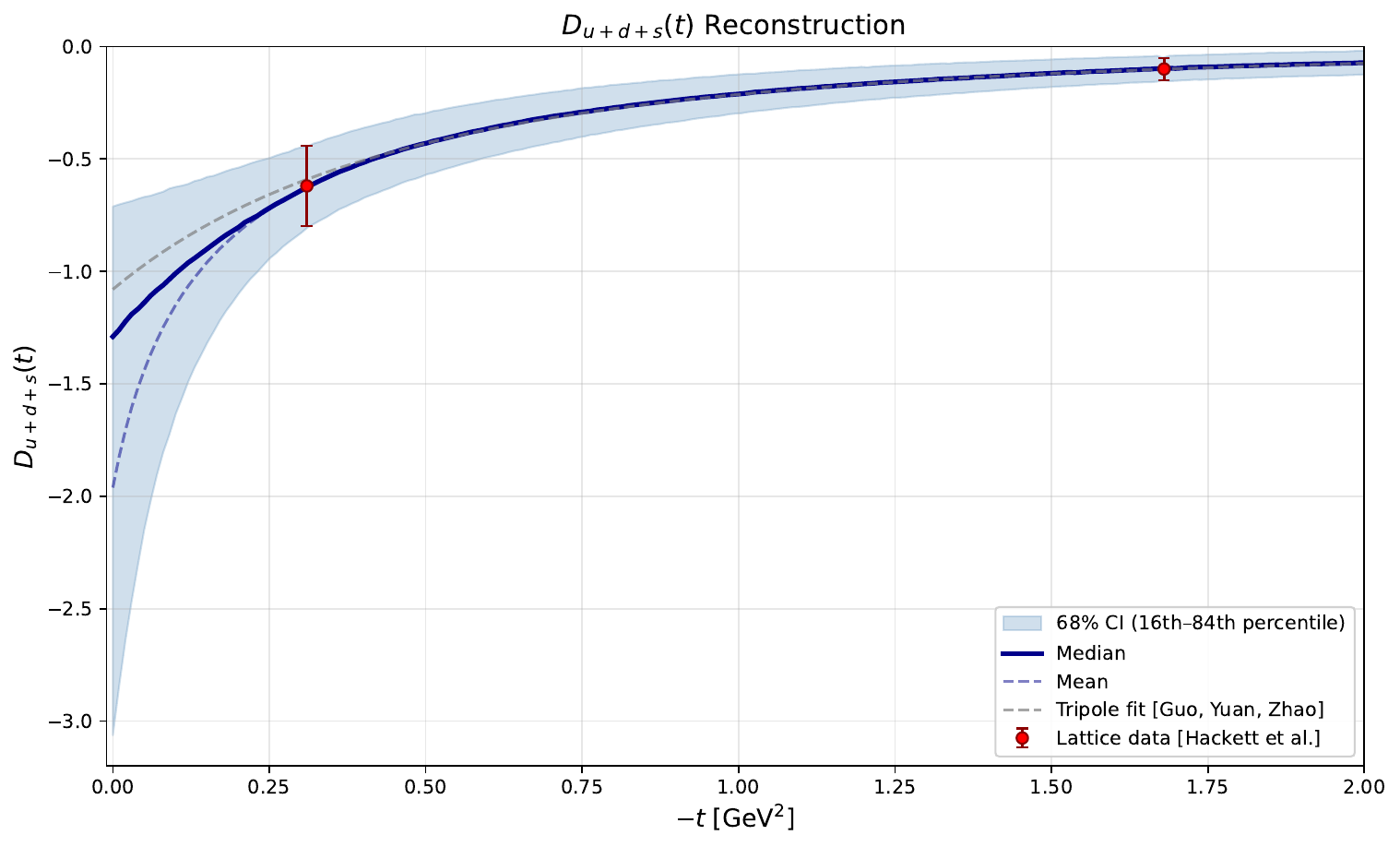}
    \caption{}
    \label{fig:D_Lattice_uds}
  \end{subfigure}
    \begin{subfigure}[H]{0.48\textwidth}
    \centering
    \includegraphics[width=\textwidth]{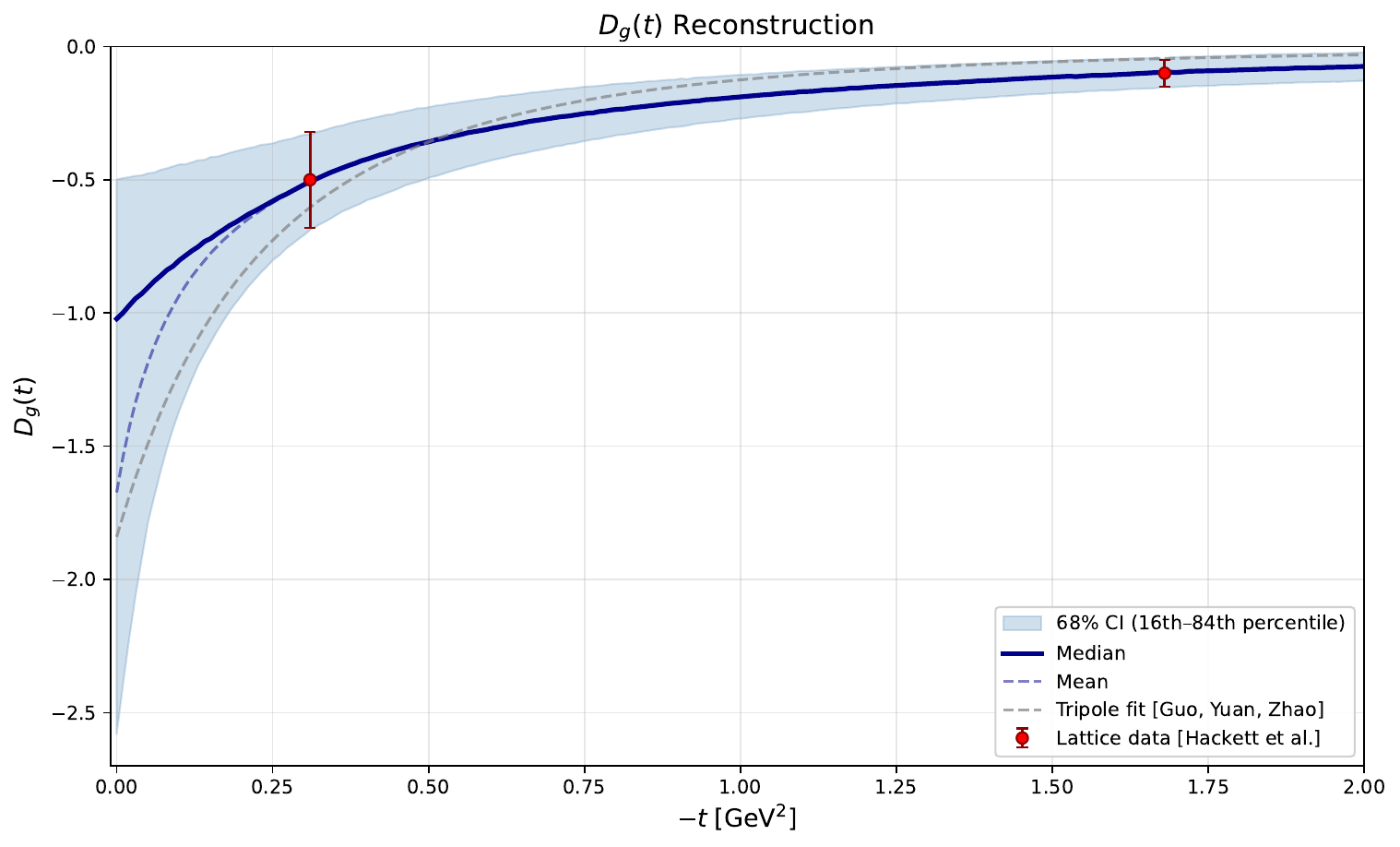}
    \caption{}
    \label{fig:D_Lattice_g}
  \end{subfigure}
  \caption{Diffusion-model reconstruction of $D(t)$ at the unphysical pion mass, conditioned on the lattice data from Ref.~\cite{Hackett:2023nkr}, compared with the model prediction from Ref.~\cite{Guo:2025jiz}.}
  \label{fig:GFF_DVCS_physical}
\end{figure}

As a further application of our approach, we reconstruct the quark and gluon contributions to the $D(t)$ form factor using two lattice data points from Ref.~\cite{Hackett:2023nkr} and compare our result with the model prediction of Ref.~\cite{Guo:2025jiz}.~As can be seen from Figs.~(\ref{fig:D_Lattice_uds}) and (\ref{fig:D_Lattice_g}), the model-independent reconstruction from only two lattice data points is consistent with the tripole parameterization of Ref.~\cite{Guo:2025jiz}.~This is again noteworthy because our method does not assume a specific functional form of the GFF.

 The reconstructions obtained above provide a continuous and parametrization-independent estimate of the proton GFFs together with their full covariance structure across the low-$|t|$ region where ChPT is applicable.~This makes it possible to extract the leading unknown LECs of the effective chiral Lagrangian directly from the diffusion-model posterior, by evaluating the ChPT expressions at the unphysical pion mass of the lattice ensemble, and without relying on any specific parametric fit.
\subsection{Extracting the unknown LECs $c_8$ and $c_9$}
\label{sec:LEC_extraction}
The low-energy constant $c_9$ is extracted by matching the diffusion-model reconstruction of $A(t)$ to our $\mathcal{O}(p^4)$ ChPT expression $A(t) = a(t) + c_9\,b(t)$ as derived in Refs.~\cite{Alharazin:2020yjv,Alharazin:2023uhr}, using generalised least squares (GLS) at five points in
$-t \in [0.02,\,0.10]\;\text{GeV}^2$, where the chiral truncation error remains below $1.3\%$.~Using the diffusion-model covariance as the GLS weight matrix, we obtain
\begin{equation}
c_9 = -0.61\,^{+0.17}_{-0.17}\;\text{(stat.)}
\;\pm\, 0.05\;\text{(ChPT trunc.)}
\;\pm\, 0.06\;\text{(fit range)}\ \rm GeV^{-1}\,.
\end{equation}
A cross-check using $J(t)$ at $\mathcal{O}(p^3)$ yields $c_9\big|_J = -0.40\,^{+0.41}_{-0.41}\ \rm GeV^{-1}$, consistent within uncertainties.~For the physical-pion-mass analysis we combine the uncertainties in quadrature and adopt $c_9 = -0.61 \pm 0.19\ \rm GeV^{-1}$.~Similarly, $c_8$ is extracted from $D(t)$ at $\mathcal{O}(p^2)$ via $D(t) = a(t) + c_8\,b(t)$ at four points in $-t \in [0.02,\,0.08]\;\text{GeV}^2$:
\begin{equation}
c_8 = -4.6\,^{+0.6}_{-0.6}\;\text{(stat.)}
\;\pm\, 0.3\;\text{(ChPT trunc.)}
\;\pm\, 0.4\;\text{(fit range)}\ \rm GeV^{-1}\,.
\end{equation}
This yields $c_8=-4.6\pm0.8\ \rm GeV^{-1}$ (all uncertainties added in quadrature).~The larger reconstruction uncertainty near $t = 0$ ($\sigma_{\mathrm{diff}} \sim 0.8$--$1.4$) reflects the absence of a sum-rule constraint on $D(0)$ and the large errors existing in the Lattice QCD at the unphysical pion mass around $t \approx 0$.~A sharper determination of $D(t)$ with the present lattice data would require including the full set of generally covariant chiral Lagrangian contributions up to $\mathcal{O}(p^4)$, incorporating explicit $\Delta$-resonance degrees of freedom, and performing a dedicated analysis to determine the additional low-energy constants that enter at this order \cite{Alharazin:2023uhr}.~

It is instructive to compare these values with the model-independent determination by Cao et al.~\cite{Cao:2025dkv}, who employ a data-driven dispersive framework in which the nucleon GFF spectral functions are constructed from $\pi\pi$ and $K\bar{K}$ intermediate states using Roy-Steiner equation solutions for the $\pi\pi/K\bar{K}\to N\bar{N}$ partial-wave amplitudes, matched to the ChPT expressions with external gravitational source.~They obtain $c_8 = -4.28\,^{+0.37}_{-0.38}~\mathrm{GeV}^{-1}$ and $c_9 = -0.68\,^{+0.06}_{-0.05}~\mathrm{GeV}^{-1}$~\cite{Cao:2025dkv}.~Both values are in excellent agreement with our extractions, $c_8 = -4.6\pm 0.8\ \rm GeV^{-1}$ and $c_9 = -0.61\pm 0.19\ \rm GeV^{-1}$, despite the two approaches being methodologically entirely distinct:~theirs reconstructs spectral functions from scattering data at the physical pion mass, while ours conditions a diffusion-model posterior on lattice QCD results at an unphysical pion mass.~That the central values agree well within one standard deviation for both LECs provides a non-trivial cross-check for both determinations.~In the next section, the extracted values of $c_8$ and $c_9$ are substituted into the ChPT expressions for the nucleon GFFs, which are then evaluated at the physical pion mass to provide approximate estimates, with uncertainties reflecting the limited high-$|t|$ information.

\section{The proton GFFs at physical pion mass}
\label{sec:GFFs}

\subsection{Reconstruction of GFFs from ChPT}
\label{sec:GFFs_ChPT}

The low-energy constants $c_8$ and $c_9$, extracted by matching the $\mathcal{O}(p^2)$--$\mathcal{O}(p^4)$ ChPT expressions for $A(t)$, $J(t)$, and $D(t)$ to the diffusion-model reconstruction at the lattice pion mass (see Sec.~\ref{sec:LEC_extraction}), are used to evaluate the proton GFFs at the physical point ($m_\pi \approx 139$~MeV).~The strategy proceeds in two steps.~First, the ChPT expressions evaluated at physical pion mass with the extracted LECs provide conditioning data concentrated in the low-$|t|$ region where the chiral expansion still has a converging behaviour.~Second, these points, together with the Poincar\'{e} constraints of Eq.~\eqref{eq:sumrules} for $A(t)$ and $J(t)$, are fed into the diffusion model to reconstruct the full $t$-dependence over the range $0 \le -t \le 2~\mathrm{GeV}^2$.

\begin{figure}[H]
  \centering
  \begin{subfigure}[H]{0.48\textwidth}
    \centering
    \includegraphics[width=\textwidth]{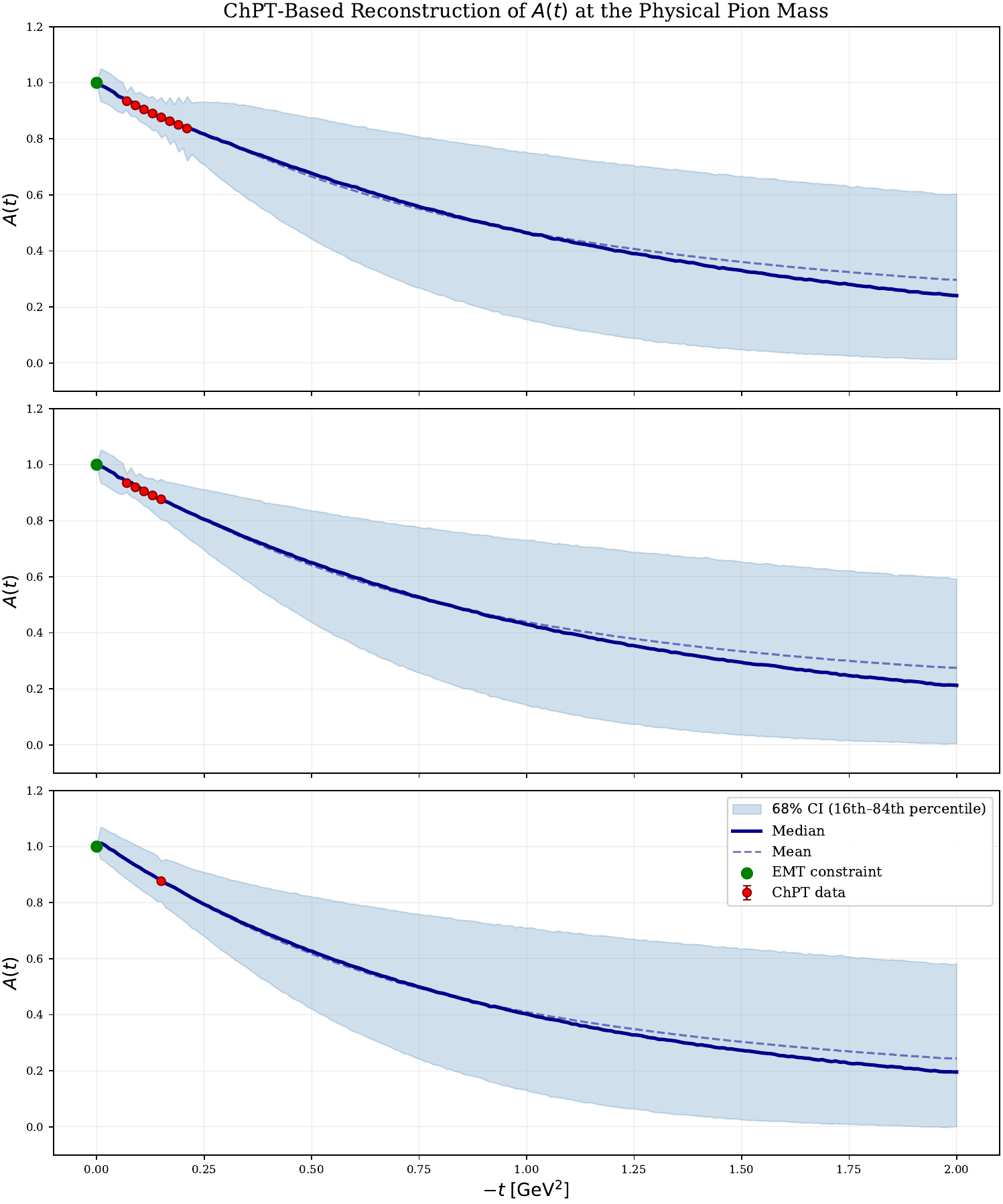}
    \caption{}
    \label{fig:A2_physical}
  \end{subfigure}
  \hfill
  \begin{subfigure}[H]{0.48\textwidth}
    \centering
    \includegraphics[width=\textwidth]{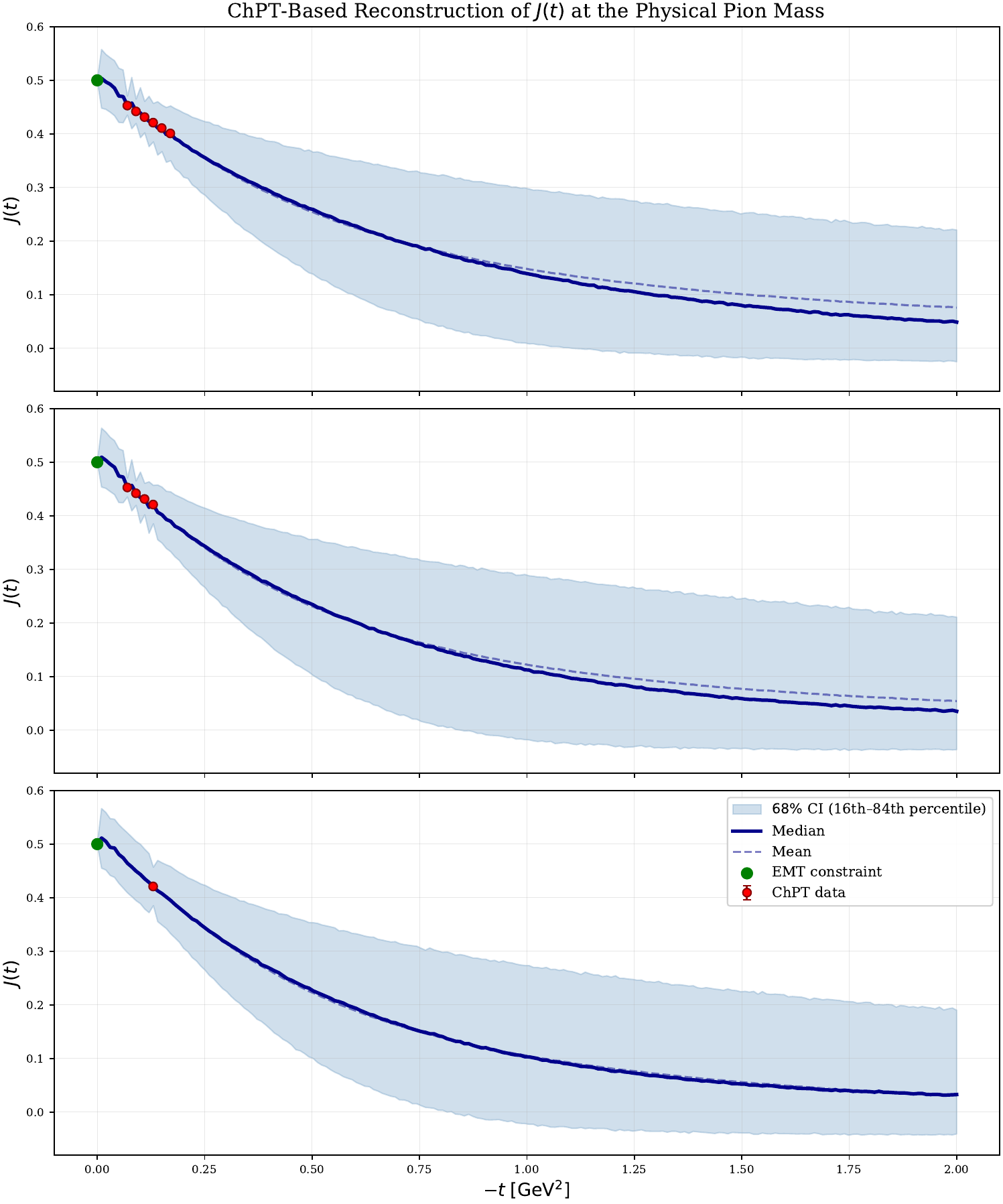}
    \caption{}
    \label{fig:J2_physical}
  \end{subfigure}
 \caption{Diffusion-model reconstruction of $A(t)$ and $J(t)$ at the physical pion mass, conditioned on ChPT data points.~Panel layout follows the data-ablation protocol, with fewer conditioning points from top to bottom}
\label{fig:GFF_physical_AJ}
\end{figure}

\begin{figure}[H]
  \centering
  \begin{subfigure}[H]{0.48\textwidth}
    \centering
    \includegraphics[width=\textwidth]{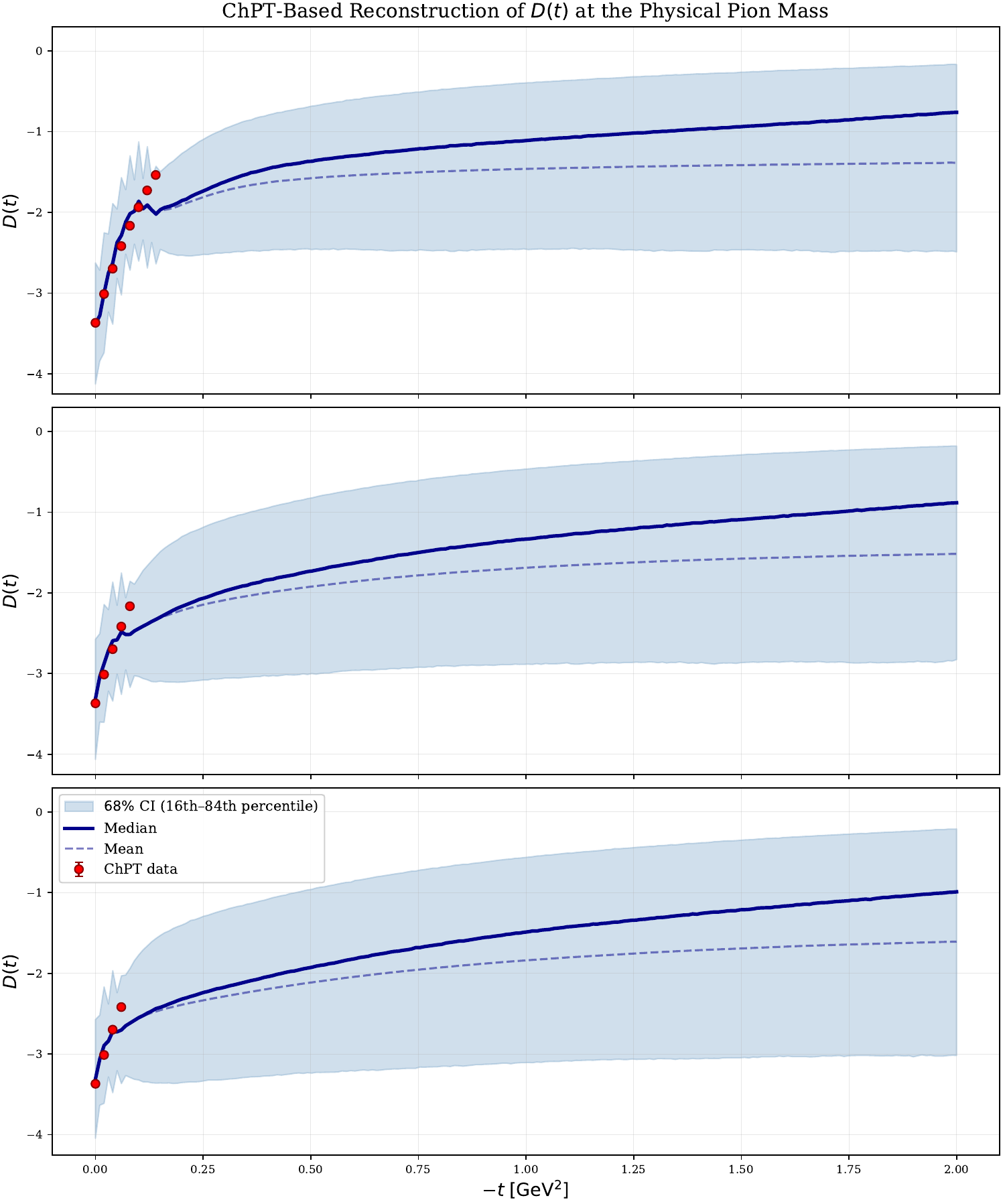}
    \caption{}
    \label{fig:D_physical}
  \end{subfigure}
 \caption{Diffusion-model reconstruction of $D(t)$ at the physical pion mass, conditioned on ChPT data points.~Panel layout follows the data-ablation protocol, with fewer conditioning points from top to bottom}
 \label{fig:GFF_physical_D}
\end{figure}

Figures (\ref{fig:GFF_physical_AJ}) and (\ref{fig:GFF_physical_D}) presents the reconstruction of all three GFFs at the physical pion mass.~For each form factor, the three panels show the effect of progressively reducing the number of ChPT conditioning points, from the full set (top) to a minimal subset (bottom).~For $A(t)$ and $J(t)$ (Figs.~(\ref{fig:A2_physical}) and (\ref{fig:J2_physical})), the Poincar\'{e} constraints at $t = 0$ anchor the normalisation, and the reconstruction remains stable even as conditioning points are removed.

In all three form factors the credible band widens appreciably at larger $|t|$.~This behaviour is expected and can be traced to two factors.~First, the most reliable conditioning data, i.e.\ those with the smallest uncertainties, are confined to the region $-t \lesssim 0.15~\mathrm{GeV}^2$; beyond this range the uncertainties on the conditioning points grow rapidly and their constraining power diminishes.~Second, the cluster of points below $-t \approx 0.15~\mathrm{GeV}^2$ alone does not provide sufficient leverage for the model to extrapolate into the intermediate-$|t|$ region with high confidence.~Indeed, the ablation study presented in the previous section (\ref{sec:Ablation}) indicates that at least one additional well-constrained point at $-t \gtrsim 0.25~\mathrm{GeV}^2$ would be required to significantly reduce the reconstruction uncertainty at intermediate momentum transfers.

The form factor $D(t)$ (Fig.~(\ref{fig:D_physical})) presents a qualitatively different situation.~Unlike $A(t)$ and $J(t)$, no sum-rule constraint fixes $D(0)$, so the reconstruction relies entirely on the ChPT conditioning data.~Moreover, the available ChPT expression for $D(t)$ is of lower chiral order than those for $A(t)$ and $J(t)$, so that the higher-$|t|$ conditioning points probe a regime where the truncated chiral expansion is no longer reliable and the quoted uncertainties underestimate the true error.~The data-ablation sequence reveals that removing these higher-$|t|$ points actually \emph{improves} the reconstruction:~the diffusion-model median is not pulled toward the potentially unreliable high-$|t|$ ChPT values but instead follows a trajectory shaped by the training prior, effectively filtering out the incomplete chiral information.

While the reconstruction of the full $t$-dependence of $D(t)$ is sensitive to the limitations discussed above, the value at $t = 0$ can be obtained directly from the extracted LEC without relying on the diffusion-model extrapolation.~Taken the physical nucleon mass $m_N = 0.938$~GeV gives us the following value of the D-term of the nucleon:
\begin{equation}
  D(0)\;=\;c_8\,m_N\;=\;-4.3\;\pm\;0.8\,,
  \label{eq:D0_prediction}
\end{equation}
where the uncertainty is propagated linearly from $\delta c_8$.~This value is consistent with the dipole-fit result $D(0)=-3.87\pm0.97$ and the $z$-expansion result $D(0)=-3.35\pm0.58$ reported by Ref.~\cite{Hackett:2023nkr} at the unphysical pion mass, with the value $D(0) = -3.38\,^{+0.34}_{-0.35}$ obtained in Ref.~\cite{Cao:2025dkv} using the dispersive techniques.\footnote{It is interesting to notice that the quark contribution alone to the D-term reported in Ref.~\cite{Goharipour:2025lep}, $D^Q(0) = -3.37 \pm 0.17$, obtained  from a $\chi^2$ analysis of Compton form factor data using skewness-dependent GPDs based on the double-distribution representation, is still compatible with our result.}~It also comfortably satisfies the chiral perturbation theory upper bound $D(0)/m_N \le -1.1(1)~\mathrm{GeV}^{-1}$~\cite{Gegelia:2021wnj}.

It is worth noting that for the form factor $D(t)$ plotted in Fig.~(\ref{fig:D_physical}), we employed the unexpanded ChPT results of Refs.~\cite{Alharazin:2020yjv,Alharazin:2023uhr}.~However, these results do not yet include all contributions to $D(t)$ arising from the third- and fourth-order chiral Lagrangians.~In particular, the fourth-order Lagrangian in covariant form has yet to be constructed in its most minimal and general form.~For the determination of the $D$-term, we therefore restricted ourselves to the low-energy region and retained only the lowest-order terms from the small-quantity expansion of the results in Refs.~\cite{Alharazin:2020yjv,Alharazin:2023uhr}.~This also explains the difference between Fig.~(\ref{fig:GFF_DVCS_physical}), where we used the complete expanded ChPT results at lowest order, and Fig.~(\ref{fig:D_physical}), where we used the unexpanded but incomplete ChPT contributions.

\subsection{Reconstruction of GFFs from DVCS data}

In Ref.~\cite{Burkert:2021ith}, the quark part of $D(t)$ form factor was extracted from deeply virtual Compton scattering data measured with the CLAS detector at Jefferson Lab.~It should be noted, however, that this extraction relies on specific parameterizations of the Compton Form Factors and the assumption of GPD H dominance, rendering the resulting $D(t)$ values inherently model dependent~\cite{Kumericki:2019ddg, Dutrieux:2021nlz}.~Nevertheless, these data points constitute the only currently available experimental determination of the gravitational form factor $D(t)$ of the proton.~In the following, we reconstruct the quark part of $D(t)$ form factor from the extracted values reported in Ref.~\cite{Burkert:2021ith} in two scenarios: first, using the data points alone, and second, incorporating the forward-limit constraint obtained in the previous section in order to illustrate the constraining power that a known forward limit provides, noting that this value refers to the total form factor and not to the quark contribution alone.~The results of the two reconstruction scenarios are shown in Fig.~(\ref{fig:GFF_DVCS_physical}).~When only the DVCS data points are used the extrapolation toward the forward limit is largely unconstrained, as reflected by the rapidly widening 68\% confidence interval and the significant deviation between mean and median at low $-t$.~Incorporating the leading-order ChPT constraint dramatically reduces the uncertainty in the forward region.

\begin{figure}[H]
  \centering
  \begin{subfigure}[H]{0.48\textwidth}
    \centering
    \includegraphics[width=\textwidth]{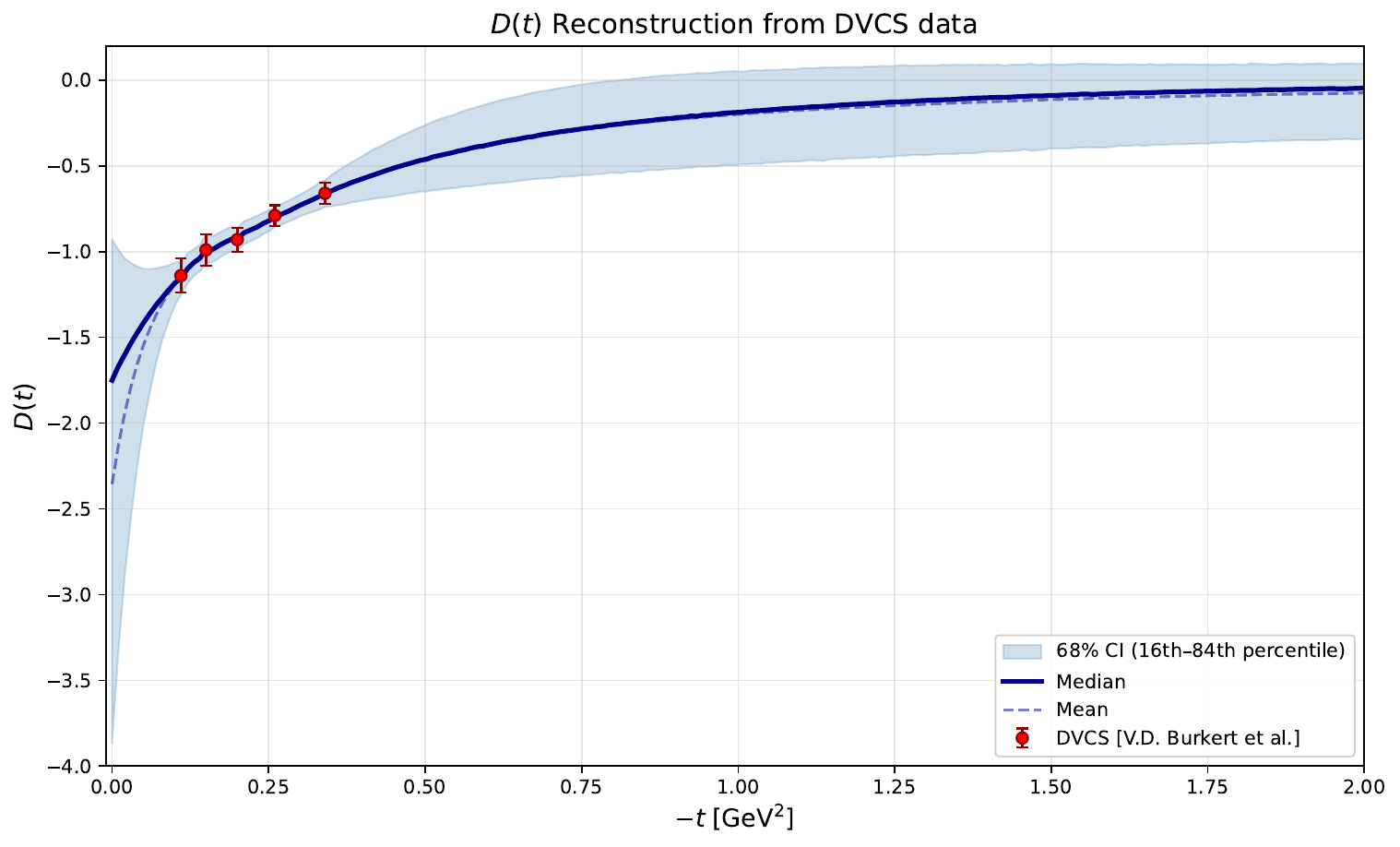}
    \caption{}
  \end{subfigure}
    \begin{subfigure}[H]{0.48\textwidth}
    \centering
    \includegraphics[width=\textwidth]{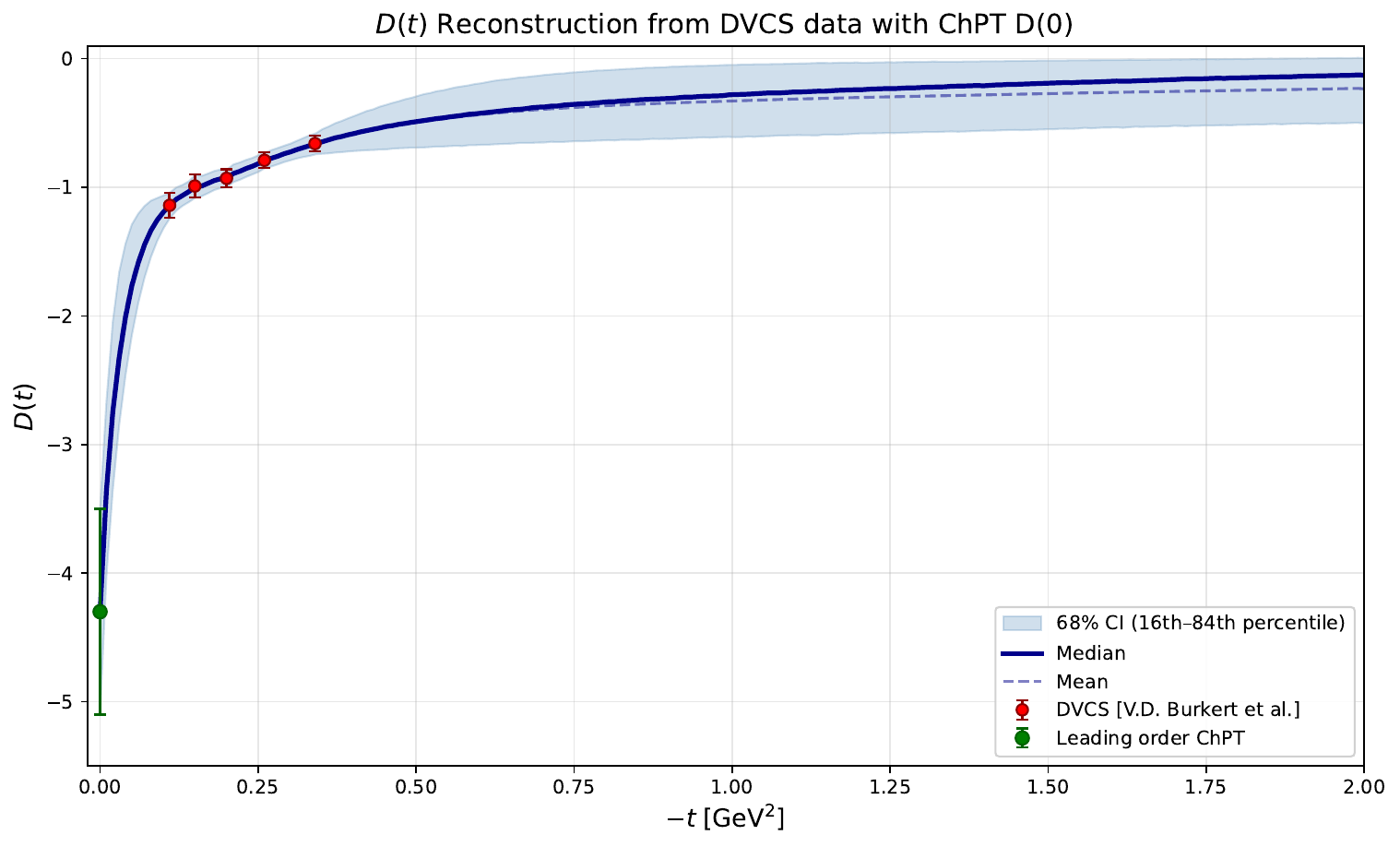}
    \caption{}
  \end{subfigure}
  \caption{Diffusion-model reconstruction of quark part of $D(t)$ at the physical pion mass, conditioned only on the DVCS data from \cite{Burkert:2021ith} and combined with the ChPT forward limit obtained in Sec.~\ref{sec:GFFs_ChPT}.}
  \label{fig:GFF_DVCS_physical}
\end{figure}

\section{Summary and Outlook}
\label{sec:Outlook}

We have developed a generative framework based on denoising diffusion for the model-independent reconstruction of hadronic form factors from sparse and noisy data.~The training prior, built from ten distinct functional classes rooted in different theoretical approaches to hadron structure, encodes a broad manifold of physically plausible shapes without privileging any single parametrization.~Applied to the proton gravitational form factors $A(t)$, $J(t)$, and $D(t)$, the framework yields non-parametric reconstructions consistent with lattice QCD and with standard parametric fits across the full kinematic range $0\le -t\le 2~\mathrm{GeV}^{2}$.~A systematic data-ablation protocol demonstrates that the reconstruction remains robust even when only one or two conditioning points are retained, confirming the strong constraining power of the physically informed prior.~The densely sampled output of the diffusion model enables a direct extraction of the chiral low-energy constants $c_8=-4.6\pm 0.8\ \rm GeV^{-1}$ and $c_9=-0.61\pm 0.19\ \rm GeV^{-1}$, in agreement with the independent dispersive determination of Ref.~\cite{Cao:2025dkv}, and yields the nucleon $D$-term $D(0)=-4.3\pm 0.8$ at the physical pion mass.

The precision of the present extraction is limited primarily by the use of the ChPT expression for $D(t)$ at $\mathcal{O}(p^2)$.~A natural next step is the construction of the complete generally covariant chiral Lagrangian up to and including $\mathcal{O}(p^4)$ with explicit $\Delta(1232)$ degrees of freedom~\cite{Alharazin:2023uhr}.~Matching the diffusion-model reconstruction to these higher-order expressions is expected to sharpen the determination of $c_8$ and $c_9$, reduce the uncertainty on the physical-mass extrapolation, and, in addition, provide access to the unknown LECs ($h_i$) of the $\Delta$-resonance Lagrangian appearing in Ref.~\cite{Alharazin:2022wjj}.~This would yield the first model-independent, ChPT-based estimates of the GFFs and $D$-term of the $\Delta$ baryon.

As already mentioned, this approach can be applied to reconstruct the GFFs of various hadrons, such as $\Delta$-resonances, $\rho$-mesons, or even nuclear systems, including electromagnetic and transition form factors.~Furthermore, our framework is naturally extensible to the reconstruction of GPDs, where the generative prior can be built from synthetic ensembles constructed via Double Distributions, ensuring that fundamental constraints such as polynomiality and positivity are satisfied by construction.~Experimental Compton Form Factors (CFFs), extracted from DVCS measurements, then serve as the conditioning data, while GFF sum rules provide additional moment constraints to further regularize the reconstruction.~Work along these lines is currently underway.

\begin{acknowledgments}
The authors thank Jambul Gegelia and Davide Laudicina for useful discussions.~We are also grateful to Xiong-Hui Cao for clarifying the chiral order at which the low-energy constants in Ref.~\cite{Cao:2025dkv} were determined, and thank Xiangdong Ji  and Latifa Elouadrhiri for helpful comments.~The calculations of this work were performed on the HPC cluster Elysium of the Ruhr University Bochum, subsidised by the DFG (INST 213/1055-1).
\end{acknowledgments}
\section*{Research Data and Code Access}
All code, training data, and analysis scripts needed to reproduce the results of this work are publicly available at Ref.~\cite{GitHub}.

\appendix
\section{Network architecture}
\label{app:network}
The denoising network $f_\theta$ is a one-dimensional ResNet-Attention hybrid that operates at constant spatial resolution~$L{=}200$ throughout, with no downsampling or upsampling layers.~This design choice reflects the fact that all grid points carry comparable physical significance and avoids information loss from pooling.~The network has the following properties:
\begin{itemize}[leftmargin=0pt]
\item It receives as input three channels $(x_\tau,\,m,\,c)\in\mathbb{R}^{3\times L}$ that are mapped to a hidden representation $h\in\mathbb{R}^{C\times L}$ ($C{=}256$) by a single convolutional layer with kernel size $K= 7$.~A symmetric output head, GroupNorm, SiLU activation, and a $1{\times}7$ convolution, projects back to a single channel, yielding the predicted $\hat v_\tau\in\mathbb{R}^{L}$.

\item The network comprises $N_{\rm res}{=}12$ residual blocks organised into $G{=}3$ groups of four, each group followed by a multi-head self-attention layer ($H{=}4$ heads).~Every residual block consists of two $1{\times}7$ convolutions interleaved with GroupNorm and SiLU activations, with a dropout rate of $0.1$ before the second convolution.~The diffusion timestep~$\tau$ is injected via Feature-wise Linear Modulation (FiLM)~\cite{Perez:2018film}:~a sinusoidal positional encoding of~$\tau$ is passed through a two-layer MLP to produce per-channel scale and shift parameters applied after the first convolution.~The self-attention layers provide a global receptive field, enabling the network to capture long-range correlations across the full grid, essential when enforcing physical constraints.

\item The second convolution in each residual block and the output projection of each attention layer are zero-initialised, so that every block acts as an identity map at the start of training.~This stabilises the initial gradient flow through the deep residual stack~\cite{Goyal:2017accurate}.
\end{itemize}
The resulting network has approximately $19\times10^6$ trainable parameters.

\section{Network training}
\label{app:training}
\begin{figure}[H]
  \centering
  \begin{subfigure}[t]{0.48\textwidth}
    \centering
    \includegraphics[width=\textwidth]{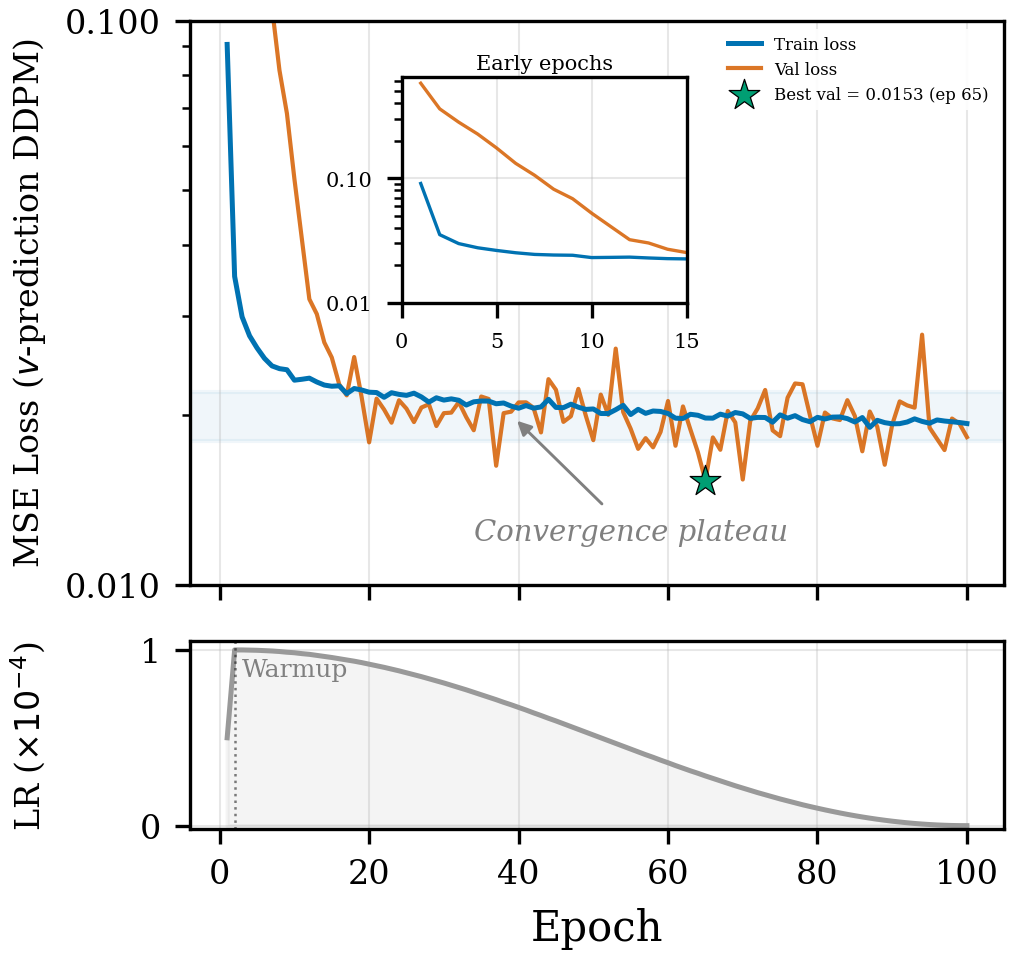}
    \caption{Training and validation loss (upper panel, log scale) and learning-rate schedule (lower panel).~The inset magnifies the first 15 epochs; the star marks the best EMA validation loss.}
    \label{fig:Training}
  \end{subfigure}
  \hfill
  \begin{subfigure}[t]{0.48\textwidth}
    \centering
    \includegraphics[width=\textwidth]{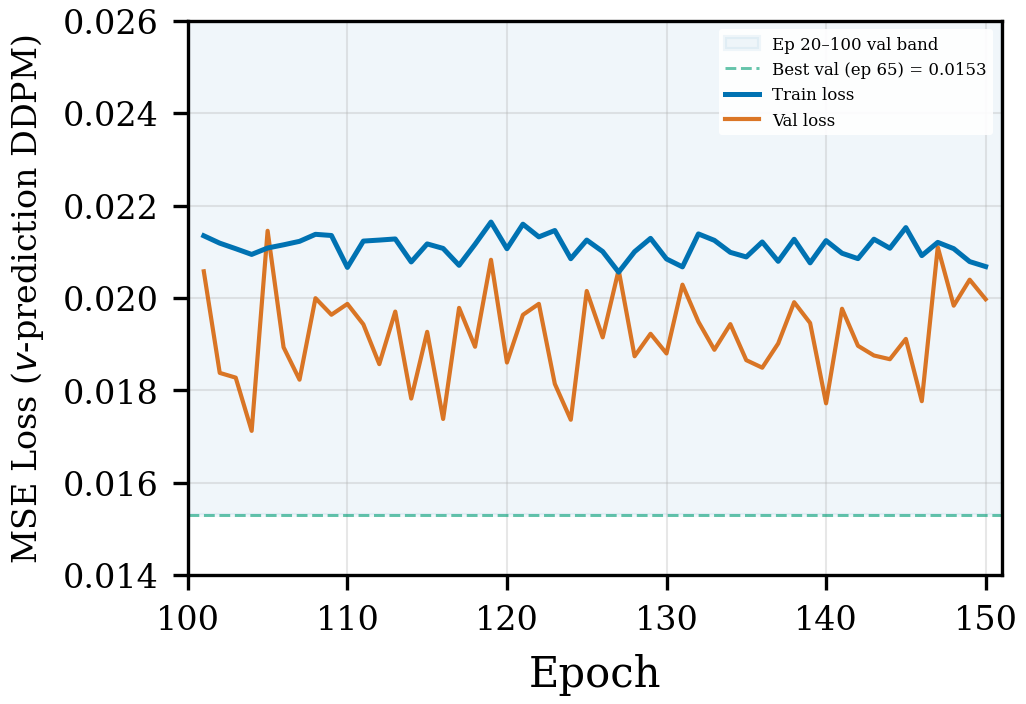}
    \caption{Fine-tuning with an expanded five-type mask distribution.~The shaded band and dashed line show the validation range and best validation loss from the original training.}
    \label{fig:fine-tuning}
  \end{subfigure}
  \caption{Training diagnostics for the $v$-prediction DDPM.}
  \label{fig:training_diagnostics}
\end{figure}

As mentioned in Sec.~\ref{sec:classes}, the model was trained on $6\times10^5$ GFF curves, each sampled on a grid of $L=200$ points and normalised to zero mean and unit variance per grid point.~A separate validation set of $6\times10^3$ curves was used for model selection.~Training minimises the $v$-prediction loss of Eq.~(\ref{eq:loss}) with the AdamW optimiser~\cite{Loshchilov:2017adamw} (learning rate $\eta=10^{-4}$, weight decay $10^{-4}$) and a batch size of 256.~The learning-rate schedule consists of a linear warmup over the first 2 epochs followed by cosine decay to zero over the remaining 98 epochs, applied per optimiser step (2\,320 steps per epoch, 232\,000 total).~Gradients are clipped at unit norm to stabilise training in the presence of the attention layers.
Moreover, an exponential moving average (EMA) of the network weights with decay $\lambda=0.9999$ is maintained throughout training;~all reported validation losses and all inference results use the EMA weights exclusively.~Checkpoints are saved every 10 epochs, and the model with the lowest EMA validation loss is retained separately.~Figure~\ref{fig:training_diagnostics}(a) shows the training and validation loss curves together with the learning-rate schedule.~Both losses decrease rapidly during the first ${\sim}15$ epochs, after which the training loss settles into a smooth descent from ${\approx}\,0.022$ to ${\approx}\,0.019$.~The validation loss, evaluated with freshly drawn random masks at each epoch, exhibits larger fluctuations but remains within the $0.015$--$0.023$ band from epoch~20 onward, indicating that the model neither overfits nor underfits.~The best validation loss of $0.0153$ is reached at epoch~65, whose EMA weights are used for all inference results reported in this work.

To verify robustness against the choice of conditioning strategy, we fine-tuned the converged model for 50 additional epochs at a flat learning rate of $2\times10^{-5}$ ($5\times$ below the initial peak, to preserve the learned representations) and a broader five-type mask distribution.~In addition to the three original mask types (see Sec.~\ref{sec:diff:conditioning}), the expanded distribution introduces a high-$|t|$ cluster (5--15 points in the last 40\% of the grid) and a sparse-spread mode (8--12 quasi-uniformly spaced points across the full grid).~As shown in Fig.~(\ref{fig:fine-tuning}), the validation loss, evaluated with the original mask distribution for comparability, remains within the $0.017$--$0.021$ band throughout, consistent with the initial convergence plateau.~This confirms that the learned GFF manifold is stable and that reconstruction quality is insensitive to the specific mask distribution.~We have further verified that varying the network architecture and training strategy leaves the validation loss within the same fluctuation band, indicating that systematic errors from these design choices are subdominant to the statistical noise floor.

\section{Ablation studies}
\label{sec:Ablation}
\begin{figure}[H]
  \centering
  \includegraphics[width=0.8\textwidth]{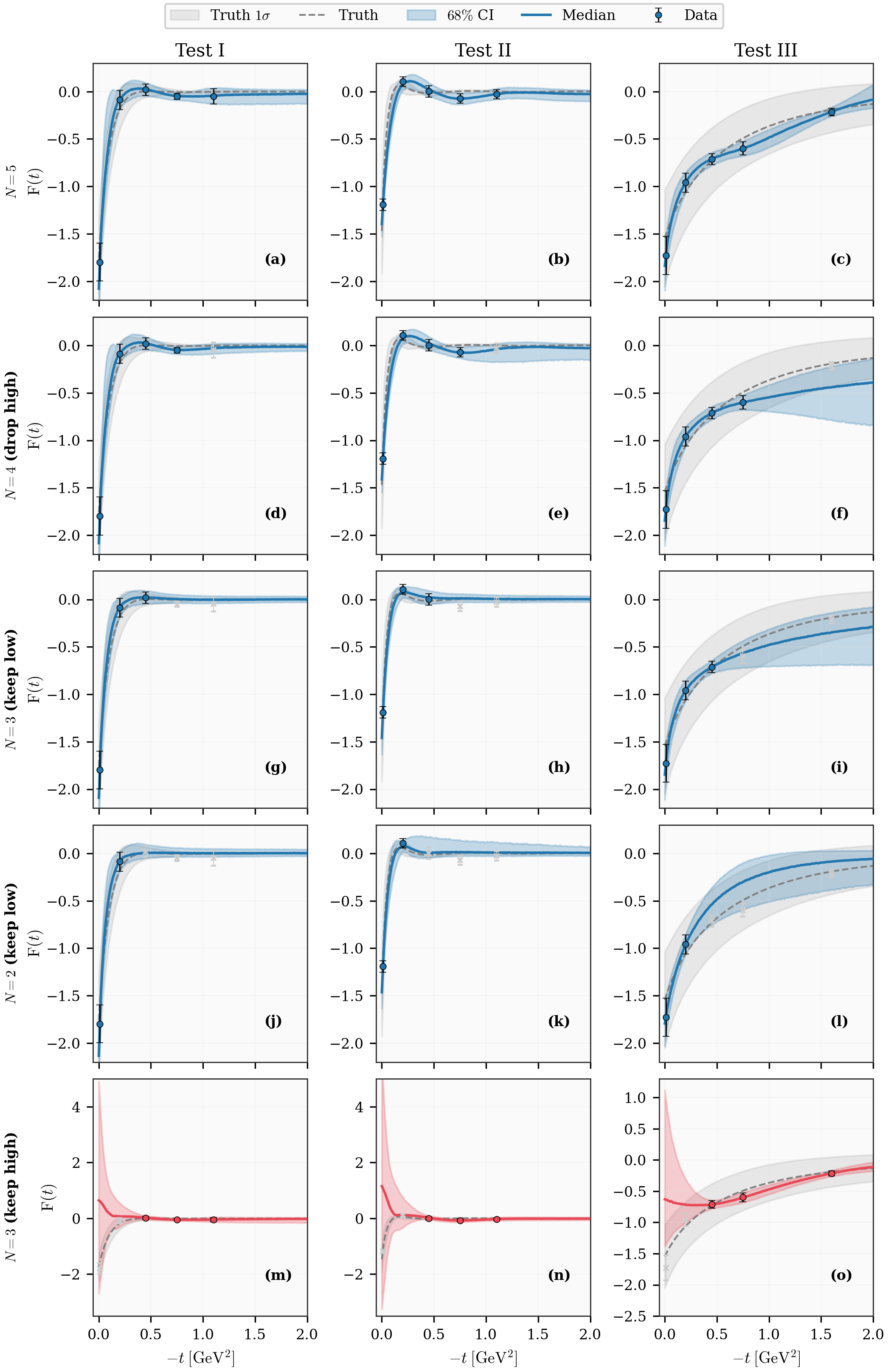}
  \caption{$N$ denotes the number of conditioning points.~Rows~1--4 progressively remove high-$|t|$ conditioning points; row~5 keeps only high-$|t|$ points for contrast.~The bottom row (red) uses the same number of points as the third row but retains only high-$|t|$ measurements, demonstrating the disproportionate importance of forward-region data.}
  \label{fig:ablation_appendix}
\end{figure}

To validate the reconstruction capability of diffusion model of this work, we perform a systematic ablation study on three synthetic test curves, shown in Fig.~(\ref{fig:ablation_appendix}).~The three scenarios are chosen to probe different form-factor behaviours:~Test curves~I and~II both exhibit a large negative value at low~$|t|$ with a sharp rise toward zero, followed by a small local maximum.~The two cases differ in the width and height of this local maximum.~Test curve~III, by contrast, displays a broad, monotonically decreasing shape with a correspondingly wider truth band.

For each scenario, a pool of candidate curves is drawn from the eight classes used in the training-data generation (Sec.~\ref{sec:classes}).~An ensemble of shape-similar curves is selected via nearest neighbours in the mean squared distance, yielding a central curve $\bar{F}(t)$ (ensemble mean) and a pointwise uncertainty $\sigma_{\mathrm{truth}}(t)$ (ensemble standard deviation).~Pseudo-data are then generated at $N$ chosen values of $-t$ by evaluating the central curve and adding Gaussian noise,
\begin{equation}
  F^{\mathrm{data}}_i
    = \bar{F}(t_i) + z_i\,\sigma_i\,,
  \qquad
  z_i \sim \mathcal{N}(0,1)\,,
\end{equation}
where $\sigma_i$ represents the experimental uncertainty at each kinematic point.~This procedure mimics realistic experimental conditions in which the true form factor is unknown and only noisy measurements at discrete momentum transfers are available.~The diffusion model is then conditioned on the tuples $\{(t_i,\, F^{\mathrm{data}}_i,\, \sigma_i)\}$ and tasked with reconstructing the full $t$-dependence, including extrapolation into unconstrained regions.~Starting from the full set of $N=5$ conditioning points, we progressively remove measurements from the high-$|t|$ end (rows~1--4 of Fig.~(\ref{fig:ablation_appendix})) and compare the resulting 68\,\% credible intervals against the known truth band.~The bottom row provides the key comparison:~three points retained at high~$|t|$ instead of low~$|t|$, with the same total number of constraints as row~3.~Across all tests, the reconstruction degrades significantly when low-$|t|$ data, which anchor the form-factor normalisation, are removed, while high-$|t|$ points can be dropped with comparatively little loss of accuracy.


\end{document}